\documentclass[apj,numberedappendix]{emulateapj}
\pdfoutput=1

\usepackage{epsfig}
\usepackage{amsmath}
\usepackage{enumerate}
\usepackage{mathrsfs}

\usepackage{natbib}
\usepackage{hyperref}

\usepackage{color}

\newbox\grsign \setbox\grsign=\hbox{$>$} \newdimen\grdimen \grdimen=\ht\grsign
\newbox\simlessbox \newbox\simgreatbox \newbox\simpropbox
\setbox\simgreatbox=\hbox{\raise.5ex\hbox{$>$}\llap
     {\lower.5ex\hbox{$\sim$}}}\ht1=\grdimen\dp1=0pt
\setbox\simlessbox=\hbox{\raise.5ex\hbox{$<$}\llap
     {\lower.5ex\hbox{$\sim$}}}\ht2=\grdimen\dp2=0pt
\setbox\simpropbox=\hbox{\raise.5ex\hbox{$\propto$}\llap
     {\lower.5ex\hbox{$\sim$}}}\ht2=\grdimen\dp2=0pt

\def\simlt{\mathrel{\copy\simlessbox}}

\def\beq{\begin{equation}}
\def\eeq{\end{equation}}

\def\sigcr{\sigma_{\rm cr}}

\def\Eq{Equation}
\def\Eqs{Equations}

\def\sigcr{\sigma_{\rm cr}}
\def\Bcr{B_{\rm cr}}

\defcitealias{2011ApJ...727L..51P}{PP}
\defcitealias{2004ApJ...609..999C}{CAZ}
\defcitealias{2010MNRAS.407L..54C}{CH}

\newcommand{\md}{\mathrm{d}}

\begin{document}

\title{
Magnetar outbursts from avalanches of Hall waves and crustal failures}

\author{Xinyu Li$^1$, Yuri Levin$^{2,1}$ and Andrei M. Beloborodov$^1$}
\affil{$^1$Physics Department and Columbia Astrophysics Laboratory, Columbia University, 
538 West 120th Street, New York, NY 10027}
\affil{$^2$School of Physics and Astronomy, Monash University, Clayton, VIC 3800, Australia}

\begin{abstract}
We explore the interaction between Hall waves and mechanical failures inside a magnetar crust, using detailed one-dimentional models that consider temperature-sensitive plastic flow, heat transport and cooling by neutrino emission, as well as the coupling of the crustal motion to the magnetosphere. We find that the dynamics is enriched and accelerated by the fast, short-wavelength Hall waves that are emitted by each failure. The waves propagate and cause failures elsewhere, triggering avalanches. We argue that these avalanches are the likely sources of
outbursts in transient magnetars. 
\medskip
\end{abstract}



\section{Introduction}

Magnetars are luminous slowly rotating neutron stars that are thought to be powered by the decay of 
ultra-strong magnetic fields $B=10^{14}-10^{16}\rm G$ (see e.g. 
\citet{2006csxs.book..547W,2008A&ARv..15..225M} for reviews). 
They have hot surfaces and produce hard X-rays flares as well as persistent nonthermal magnetospheric radiation.
So-called transient magnetars show periods of low luminosity followed by months-to-years long outbursts, increasing X-ray luminosity by 
up to 3 orders of magnitude \citep{2004ApJ...609L..21I, 2004ApJ...605..368G,2008A&ARv..15..225M,2011ASSP...21..247R}.  

The mechanism triggering this activity is not established.
Following the original ideas of \citet{1996ApJ...473..322T}, 
\citet{2011ApJ...727L..51P} 
argued that the outbursts are powered by localized releases of elastic energy in the crust  due to mechanical failures in the crystal lattice (the ``starquakes"). 
They modeled the build up of the elastic energy before each release as a result of the changing magnetic stresses, and have produced a phenomenological numerical model for the frequency of the outbursts and the magnitude of their energy release.  \citet{2012ApJ...750L...6P} have modeled the outbursts by computing the thermal flux emerging from the neutron star surface from an impulsive energy release in the crust (see \citet{2006MNRAS.371..477K,2014MNRAS.442.3484K} for a detailed discussion of heat transfer through the
 magnetar's crust). 

The magnetic field in the crust evolves due to a combined action of the Hall drift and ohmic dissipation \citep{1992ApJ...395..250G, 2004MNRAS.347.1273H,2013MNRAS.434..123V}. The multi-dimensional dynamics of the Hall drift in the neutron-star crust is complex and not fully 
understood, although significant new insights have come from recent numerical experiments
\citep{2012MNRAS.421.2722K, 2013MNRAS.434.2480G,2014MNRAS.438.1618G,2015MNRAS.446.1121G, 2015PhRvL.114s1101W}. It is generally thought to proceed on
long timescales of $\sim 10^3$ years in the deep crust.

The gradual evolution of the magnetic field due to Hall drift 
is punctuated by
shearing motions of the crust that relieve magnetic stresses. 
\citet{2014ApJ...794L..24B} proposed that these yielding motions occur through 
transient thermoplastic waves, which resemble deflagration fronts burning magnetic 
energy. These fronts leave sharp gradients in the crustal magnetic fields, which must 
feedback on the field evolution through Hall drift. The goal of the present paper is to
explore the interaction between the Hall evolution 
and the mechanical failures that such evolution induces. We show that each failure produces a burst of 
short Hall waves which speed up the evolution. The short Hall waves
propagate to different parts of the crust and cause new mechanical failures, thus producing an avalanche. 
We propose that these avalanches are the mechanism of outbursts in transient magnetars.
 
We build a detailed one-dimensional model which follows the following processes:
\newline 1. 
Rapid plastic motions driven by super-critical magnetic stresses,
and the associated emission of short Hall waves.
\newline 2. 
Transport of the generated plastic heat 
and neutrino cooling of the crust.
This allows us to find
the thermal flux emitted from the stellar surface.
\newline 3. 
Surface shear resulting from the crustal avalanches,
and the associated Poynting flux into the magnetosphere.
This flux can feed a magnetospheric activity.

The plan of the paper is as follows. 
Sections~\ref{dyn}-\ref{avalanche} describe the mechanism of Hall-wave generation, plastic failures, and 
the crustal failure development through the Hall-mediated avalanche. Section~\ref{twist} introduces 
a simplified model of magnetospheric twisting by the crustal failures that is adapted for 
one-dimensional simulations. Section~\ref{sim} presents a numerical simulation of the 10-kyr 
evolution of the crustal magnetic field coupled to the magnetosphere. Connections of the model 
to observations are discussed in Section~\ref{discussion}.


\section{Hall waves}
\label{dyn}
The magnetic field evolution in the crust is governed by the equation
\beq
\dot{\boldsymbol{\mathcal{B}}}=\nabla\times\left(\boldsymbol{v}\times\boldsymbol{\mathcal{B}}\right)+
\nabla\times\left(\eta\nabla\times\boldsymbol{\mathcal{B}}\right).
\eeq
Hereafter dot above a symbol signifies time derivative $\partial/\partial t$.
The first term on the right-hand side represents advection of the magnetic field by the 
electron fluid, which is moving with 
velocity $\boldsymbol{v}$ in the frame of undeformed crust, 
and the second term represents  
ohmic diffusion, with the diffusivity given by
\begin{equation}
\eta={c^2\over 4\pi \gamma}, 
\end{equation}
where
$\gamma$ is the electrical conductivity of the crust. The electron velocity consists of three components: \begin{equation}
\boldsymbol{v}=\boldsymbol{v}_{\rm H}+\dot{\boldsymbol{\xi}}_{\rm ela}+\dot{\boldsymbol{\xi}}_{\rm pl}.
\label{velectron}
\end{equation}
 Here $\boldsymbol{v}_H$ is the Hall drift velocity; it describes the electron motion
 relative to the ions and is related to the 
 electric current density $\boldsymbol{j}$,
\beq
\boldsymbol{v}_{H} = -\frac{\boldsymbol{j}}{n_e e}=-{c\over 4\pi n_e e}\nabla\times\boldsymbol{\mathcal{B}},
\eeq
where $n_e$ is the electron density and 
$-e$ is the electron charge. 
The other two terms on the right-hand side of Equation~(\ref{velectron})
represent the motion of the ions; $\boldsymbol{\xi}_{\rm ela}$ is the elastic deformation 
of the lattice and $\boldsymbol{\xi}_{\rm pl}$ is the plastic deformation.
The ion motion was
neglected in all studies of the Hall evolution of the crustal 
magnetic field except the work of \citet{2004ApJ...609..999C}, who discussed the 
contribution of 
$\boldsymbol{\xi}_{\rm ela}$ to the field dynamics and showed that it 
dramatically changes the dispersion relation of Hall waves in the upper layers of the crust.

Similar to \citet{2004ApJ...609..999C},
we consider a simplified plane-parallel configuration with the vertical $z$ axis pointing
from the core to the surface. 
\footnote{In our simplified 1D model the Hall evolution term is linear; the non-linearity enters into our model through the yielding of the crust to magnetically-induced stresses.}
The ion lattice displacement 
$\boldsymbol{\xi}=\boldsymbol{\xi}_{\rm ela}+\boldsymbol{\xi}_{\rm pl}$ 
is purely horizontal, and the model is one-dimensional in the sense that all variables  
(magnetic field, displacement, temperature, conductivity, etc.) are functions of $z$ and 
time $t$. Then $\boldsymbol{\mathcal{B}}(z,t)=(B_x,B_y,B_z)$ has a constant $B_z$ component, and the evolution 
equation for the horizontal field reduces to
\begin{eqnarray}\label{vel}
\dot{B}_a=B_z\partial_z \left[v_{H,a}+\dot{\xi}_a\right]+\partial_z\left(\eta\partial_z B_a\right),
\end{eqnarray}
where 
index $a=x,y$ corresponds to the horizontal components.
It is convenient to define a complex-valued magnetic field $B\equiv B_x+i B_y$, displacement $\xi\equiv\xi_x+i\xi_y$, etc. 
Then
\beq
v_H\equiv v_{H,x}+i v_{H,y} = -i\frac{c}{4\pi n_e e}\partial_z B,
\eeq
and the evolution equations reads
\beq
\label{hall}
\dot{B}= -i\partial_z\left(D\partial_z B \right)+B_z\partial_z\dot{\xi}.
\eeq
Here $D=D_H+i\eta$, 
and the Hall diffusion coefficient is
\begin{equation}
 D_{H}=\frac{B_z c}{4\pi n_e e}.
\end{equation}
 
\subsection{Generation of Hall waves}\label{generation}
In this paper we will explore two processes that can produce Hall waves in the crust. 
First, \citet{1996ApJ...473..322T} argued that a sudden rearrangement of magnetic field 
lines in the core can launch Hall waves from the core-crust interface into the crust. 
Fast magnetic rearrangement due to hydromagnetic instability during 
the early life of the
magnetar or rapid ambipolar diffusion in the hot magnetar core 
\citep{1992ApJ...395..250G,2016arXiv160509077B} 
can create a current sheet at the core-crust interface. 
Configurations with the current sheet at the interface was considered by \citet{2013PhRvL.110g1101L,2015MNRAS.453..671G} in the context of superconducting stars, and also appeared in some simulations of the core field expulsion \citep{2016MNRAS.456.4461E}.
This localized horizontal current drags the field lines and launches 
a train of Hall waves that propagate toward the top of the crust.
More importantly, local sudden mechanical failures in the crust can induce a
local change in the horizontal magnetic field. This change generates horizontal currents 
that drag the field lines and launch Hall waves which propagate both upward and downward
from the failure.

It is instructive to consider first a simple analytical example of how a burst of Hall waves is 
generated; this example was studied by \citet{2015MNRAS.447.1407L} in a different context. 
In this example the crust is a homogeneous, infinitely rigid, ideal conductor 
(these assumptions will be relaxed below). The field evolution equation then becomes
\begin{equation}
\label{eq:Hw}
\dot{B} +iD_H\partial_z^2 B=0.	
\end{equation}
This equation admits simple wave solutions with the dispersion 
relation $\omega=D_H k^2$. Note that shorter waves (large wavenumbers $k$) 
have higher phase speeds $\omega/k$ and group speeds $2\omega/k$.
The Green's function $G(z,z';t)$ for Equation~(\ref{eq:Hw}) is given by
\begin{equation}
G(z,z';t) = 
\frac{1}{\sqrt{-4\pi i D_H t}}\exp\left[-i\frac{(z-z')^2}{4 D_H t}\right].	
\end{equation}
Given an initial condition $B(z,0)$ one can find the solution,
\begin{equation}
	B(z,t) = \int\md z'\: G(z,z';t)B(z',0).
\end{equation}

\begin{figure}[t]
\vspace*{-0.3cm}
\hspace*{-0.5cm}
\begin{tabular}{c}
\includegraphics[width=0.5\textwidth]{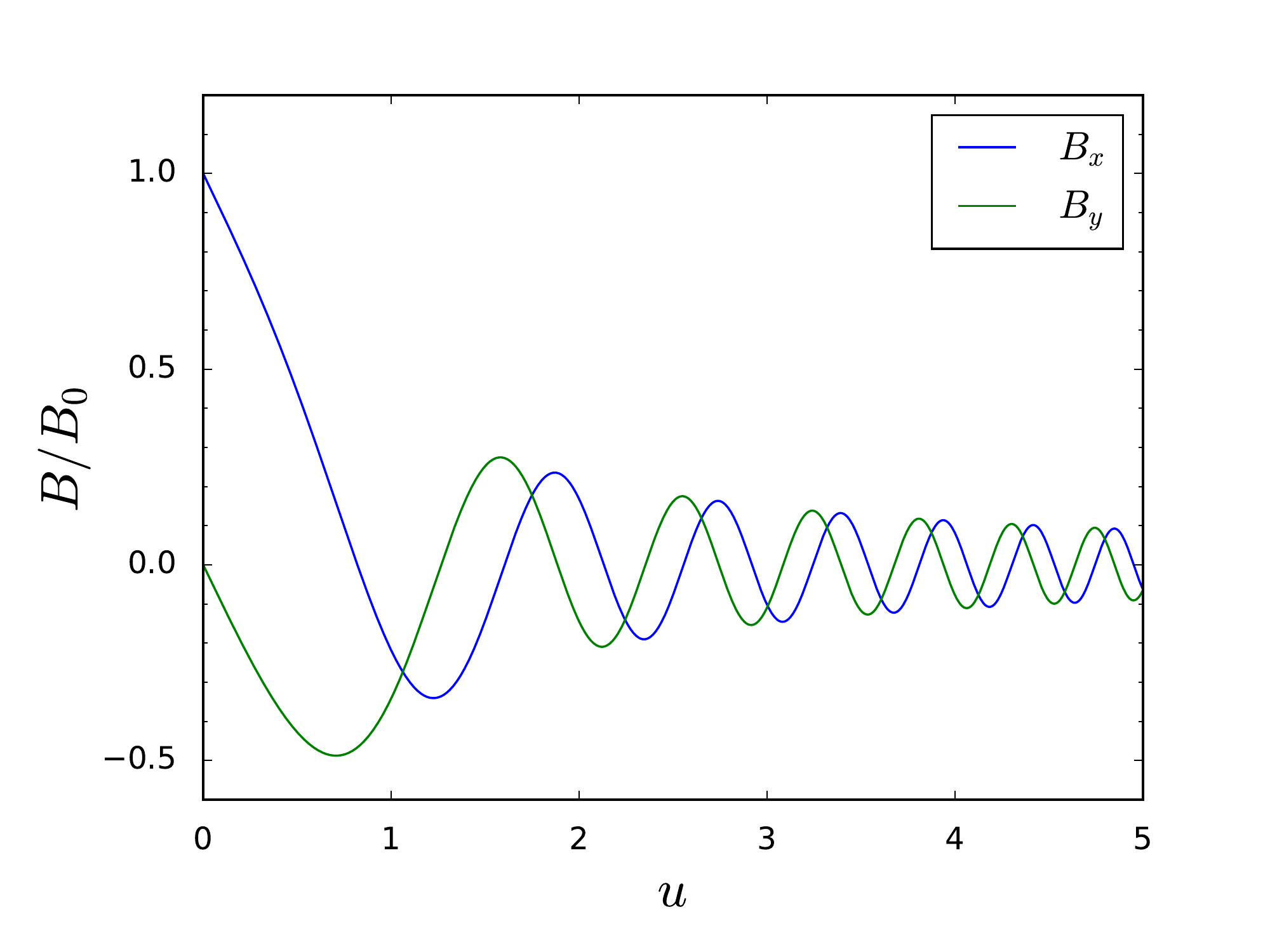} 
\end{tabular}
\vspace*{-0.3cm}
\caption{
Self-similar wave $B_x(u)$, $B_y(u)$ (where $u = z/\sqrt{2D_H t}$) in a 
homogeneous crust of infinite conductivity with the initial $B=B_x+iB_y=0$. 
The wave is launched at $t=0$ and $z=0$ by the 
jump of $B_x$ from 0 to $B_0$.
$B_x=B_0$ is kept fixed at the boundary $z=0$ while the Hall evolution of 
$B(z,t)$ washes out the jump with time. The constant profile $B(u)$
implies the self-similar stretching of $B(z)$ as $z\propto t^{1/2}$, from initially
infinitesimal to arbitrary large widths.
}
\label{fresnel1}
\end{figure}

Let us consider the initial condition $B(z,0)=0$ everywhere in the crust except
its boundary with the core, where the field jumps to $B_0\neq 0$. The boundary 
condition is fixed at $B_0$ throughout the evolution. Then the solution is given by
\begin{equation}\label{fre1}
  \frac{B(z,t)}{B_0} = 1-\sqrt{\frac{2}{\pi}}\frac{2}{1-i}\left[\mathcal{C}(u)-i\mathcal{S}(u)\right],
\end{equation}
where $u = z/\sqrt{4D_H t}$; $\mathcal{S}$ and $\mathcal{C}$ are Fresnel integrals,
\beq
	\mathcal{S}(u) = \int_0^u \sin(s^2)\:\md s,  \quad
	\mathcal{C}(u) = \int_0^u \cos(s^2)\:\md s.
\eeq
The solution is self-similar; its dependence on $u$ is shown in Figure~\ref{fresnel1}.
It demonstrates how the horizontal current sheet at the boundary generates a broad 
spectrum of Hall waves. The fast short waves lead the longer waves.

\begin{figure*}
\begin{centering}
\includegraphics[width=1.0\textwidth]{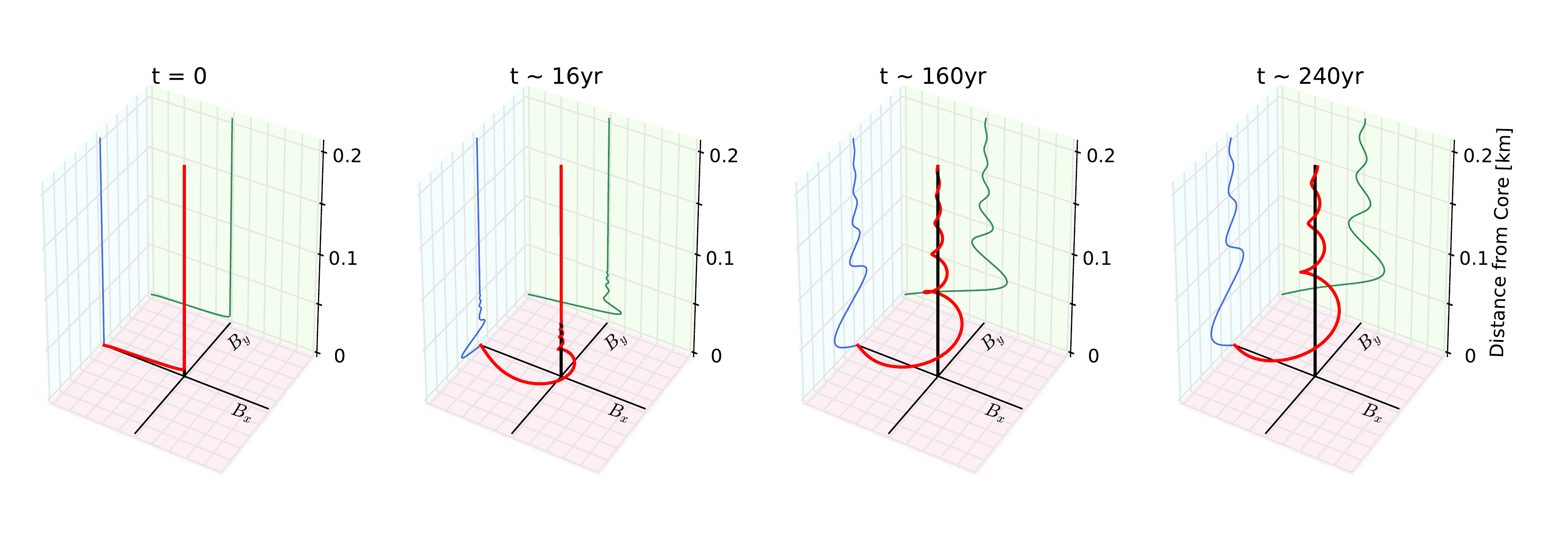} 
\end{centering}
\caption{
Generation of Hall waves by a jump of $B_x$ from $B_x=0$ at $z\gtrsim 10$~m to 
$B_x=B_0=6\times 10^{15}$~G at $z=0$ (the core-crust interface). The four snapshots show 
the evolution of $B_x$ and $B_y$; vertical field $B_z=3\times 10^{14}$~G remains constant. 
The evolution is calculated using a realistic density profile $\rho(z)$ and conductivity 
$\gamma(z)$ of the crust. The black lines indicate the $x,y,z$ axes.  The red curve traces 
the end of the horizontal vector $(B_x,B_y)$, which is a function of the vertical position $z$.
The green and blue projections show $B_x(z)$ and $B_y(z)$; they resemble the 
self-similar solution shown in Figure~\ref{fresnel1}.
}
\label{cartoon}
\end{figure*}

The same problem with a realistic density profile $\rho(z)$ and electric conductivity 
$\gamma(z)$ can be solved numerically using \Eq~(\ref{hall}) (neglecting crustal 
deformation $\xi$). The finite resistivity leads to efficient damping of the fast Hall waves with 
high wavenumbers, limiting the speed of the wavefront launched by the jump of $B_x$.
Figure~\ref{cartoon} 
shows four snapshots of the resulting evolution of the magnetic field.

Similar waves will be launched by a plastic flow that has created a jump in $B$ at 
some $z_0>0$ inside the crust, as will be discussed below. In this case, one will need to 
trace the crustal Hall waves excited at $z<z_0$ and $z>z_0$.


\section{Plastic failures}\label{plastic}

\subsection{Stress balance}

Hall evolution can generate strong shear stresses $BB_z/4\pi$, where 
$B$ is the horizontal magnetic field.
As long as the external $B=0$ at the top of the crust (this assumption will be relaxed in 
Section~5) the stress balance is only possible 
if the entire stress $BB_z/4\pi$ is offset by the elastic stress of the ion lattice,
\beq
\label{eq:stress}
   \frac{B B_z}{4\pi}=-\mu\,\partial_z \xi_{\rm ela}=\sigma,
\eeq
where $\mu$ is the shear modulus of the lattice. We expect the stress balance 
to be satisfied to a high precision at all times, even during a crustal failure when a plastic flow occurs \citep{2014ApJ...794L..24B}, provided that the plastic flow is slow compared with the relaxation to stress balance. The latter occurs on the shear-sound-crossing timescale $<$0.1~s. 

Taking the time derivative on both sides 
of \Eq~(\ref{eq:stress}) and substituting
into \Eq~(\ref{hall}), we get
 
\beq
\label{eq:evol}
\left(1+\frac{\mu_B}{\mu}\right)\dot{B}= -i\partial_z \left(D {\partial_z B} \right)+B_z\partial_z\dot{\xi}_{\rm pl},
\eeq
where $\mu_B\equiv B_z^2/4\pi$.
The above equation,
but without the plastic deformation term on the right-hand side, was derived 
and used to obtain the dispersion relation of  
Hall waves in \citet{2004ApJ...609..999C}. The elastic deformation of the crust strongly affects the Hall-wave propagation when $\mu<\mu_B$.

\subsection{Mechanical failure}\label{mechanical}

When the shear stress in the crust reaches a critical value 
\beq
\sigma_{\rm cr}\sim 0.1\mu,
\eeq
the crust must yield inelastically, as demonstrated in numerical experiments by 
\citet{2009PhRvL.102s1102H} and 
\citet{2010MNRAS.407L..54C}.
They propose that $\sigcr$ depends on temperature as
\beq\label{shearstress}
\sigma_{\rm cr}(T) = \sigma_{\rm cr}(0)\left(1-\frac{65.128}{\Gamma-71}\right), 
\eeq
where $\Gamma=Z^2 e^2/akT$ is the Coulomb coupling parameter for ions with charge
number $Z$ and separation $a=(3/4\pi n_{\rm i})^{1/3}$ ($n_{\rm i}$ is the ion number density.)
These authors studied rapidly shearing boxes of $\sim 100\times100\times 100$ lattice sites, 
and found that after the shear stress reaches $\sim 0.1\mu$ the failure develops with the 
stress reduced by an order of magnitude.
When the shear ends, the crystal strength must eventually heal.
Our model below will assume a similar behavior of the elastic stress in macroscopic
failure events, although the extrapolation of the small-scale rapid-shear experiments 
to slowly fostered macroscopic failures may not be reliable.
Note also that two-dimensional shear failures, common in the earth crust as sources of 
earthquakes, are strongly suppressed in a magnetar by magnetic tension 
\citep{2012MNRAS.427.1574L}. The failure may be described as 
a plastic flow \citep{2014ApJ...794L..24B}, see also \citet{2003ApJ...595..342J}.

The critical stress $\sigcr$ defines the critical magnetic field $\Bcr$,
\beq
  \frac{\Bcr B_z}{4\pi}=\sigcr.
\eeq
Motivated by the results of \citet{2009PhRvL.102s1102H}
we will
assume that once the failure
is initiated the critical stress  drops
to $\sigma_{\rm cr}^{\rm new}=0.1\sigma_{\rm cr}$. 
Correspondingly, $\Bcr^{\rm new}=0.1\Bcr$ during the plastic flow.
This prescription makes sure that the horizontal magnetic field evolves to the lower value $\Bcr^{\rm new}$ in a visco-elastic manner. Once the elastic stress reaches $\sigma_{\rm cr}^{\rm new}$ we assume that the lattice reforms and the visco-elastic evolution stops.
Our results are not sensitive to the exact choice of $\sigma_{\rm cr}$.

Finally, we specify the plastic flow rate (the last term in 
\Eq~(\ref{eq:evol})) using
a simple viscoelastic model,
\beq
B_z\partial_z \dot{\xi}_{\rm pl}=-\alpha B\left(1-\frac{B_{\rm cr}^{\rm new}}{|B|}\right)\Theta\left(|B|-B_{\rm cr}^{\rm new}\right).
\label{plasticprescription}
\eeq
Here $\Theta(...)$ is the Heaviside step function, so the plastic flow occurs as long as
$|B|>B_{\rm cr}^{\rm new}$. \Eq~(\ref{plasticprescription}) describes a shear failure motion 
that relaxes the local magnetic stress by driving the field $B$ from $\Bcr$ to $\Bcr^{\rm new}$.
The parameter $\alpha$ has the dimension of $s^{-1}$ and determines the relaxation rate.
After time $\tau\gg \alpha^{-1}$ from the beginning of the plastic failure, $B$ 
becomes exponentially close to $\Bcr^{\rm new}$ and the crystal should heal, i.e. the critical 
stress 
should increase
back to $\sigma_{\rm cr}\sim 0.1\mu$, ending the plastic flow. 
The qualitative results of this paper were found to be weakly affected by the choice of rate 
$\alpha$ and healing time $\tau$ for reasonably fast plastic flow rates.
In the simulations presented below we use $\alpha=10^{-4}$~s$^{-1}$
and $\tau = 1$~yr.

\subsection{Heat transfer and thermoplastic waves}

The energy density that can be dissipated through plastic failures is 
the sum of magnetic and elastic energies,
\beq
 U = \frac{|B|^2}{8\pi}+\frac{1}{2}\mu |\partial_z \xi_{\rm ela}|^2 
 = \left(1+\frac{\mu_B}{\mu}\right)\frac{|B|^2}{8\pi}.
\eeq 
The dissipation rate due to plastic flow is given by
\begin{equation}
   q_{\rm pl}={B_z\left|B\partial_z \dot{\xi}_{\rm pl}\right|\over 4\pi},
\label{qpl}
\end{equation}
where $|B|=(B_x^2+B_y^2)^{1/2}$.

\citet{2014ApJ...794L..24B} showed that the temperature-softening effect,
i.e. the reduction of $\sigcr(T)$ with increasing $T$, allows the 
plastic failure to propagate through heat diffusion.
The resulting thermoplastic wave (TPW) resembles a deflagration front, and its speed is 
\beq
\label{eq:vTPW}
  v\sim(\alpha\chi)^{1/2},  
\eeq
where $\chi=\kappa/C_V\sim 10-100$~cm$^2$~s$^{-1}$
is the heat diffusion coefficient, 
with $C_V$ and $\kappa$ being the heat capacity and thermal conductivity of
the crustal
material.
The TPWs are much faster than the Hall waves, 
so the Hall evolution is negligible during the crustal failure through a TPW.

The TPW
propagation requires that
the magnetic field ahead of the wave $B_0$ 
be sufficiently close to $\Bcr$, so that heat diffusion from the plastic flow 
is capable of reducing
$\Bcr(T)$
below $B_0$. A simple propagating solution $B(z-vt)$ is obtained from 
\Eq~(\ref{plasticprescription}) assuming a uniform medium ahead of the wave with 
a uniform field $B_0$. Suppose the wave was launched at $z_\star$ at time $t_\star$.
In the plastically flowing region $z-vt<z_\star-vt_\star$,
\Eq~(\ref{plasticprescription}) gives
\beq
-v \frac{dB}{dw} = -\alpha B\left(1-\frac{\Bcr^{\rm new}}{|B|}\right), 
\qquad w=z-vt.
\eeq
Then one finds the solution,
\beq\label{tpw}
B = (B_0-B_1)\exp\left[\frac{\alpha}{v}(w-w_\star)\right]+B_0,  \quad w<w_\star,
\eeq 
which describes the plastic relaxation of $B_0$ to a weaker field $|B_1|=\Bcr^{\rm new}$.
One can see that the characteristic thickness of the wave front is $v/\alpha$.

A TPW launched in a non-uniform background will eventually extinguish, leaving 
a jump of the magnetic field $B_0\rightarrow B_1$ of width $\sim v/\alpha$. This jump
affects the subsequent Hall evolution of the magnetic field.


\section{Hall-mediated avalanche}\label{avalanche}

The jumps of the horizontal magnetic field as a result of plastic failures generate Hall waves.
This can be illustrated 
by an idealized model of a homogeneous crust with an initially uniform field $B_0$ that was 
suddenly changed to $B_1=0.1B_0$ at $z<z_0$. 
$|B_1|=\Bcr^{\rm new}$ carries the same meaning as in Equation (\ref{tpw}).
The problem is similar to that considered in 
Section~\ref{dyn} except that now $B$ jumps from $B_0$ to $B_1$ at $z_0$ instead of jumping 
from $B_0$ to $0$ at $z=0$.

\begin{figure}[t]
\vspace*{-0.3cm}
\hspace*{-0.5cm}
\begin{tabular}{c}
\includegraphics[width=0.52\textwidth]{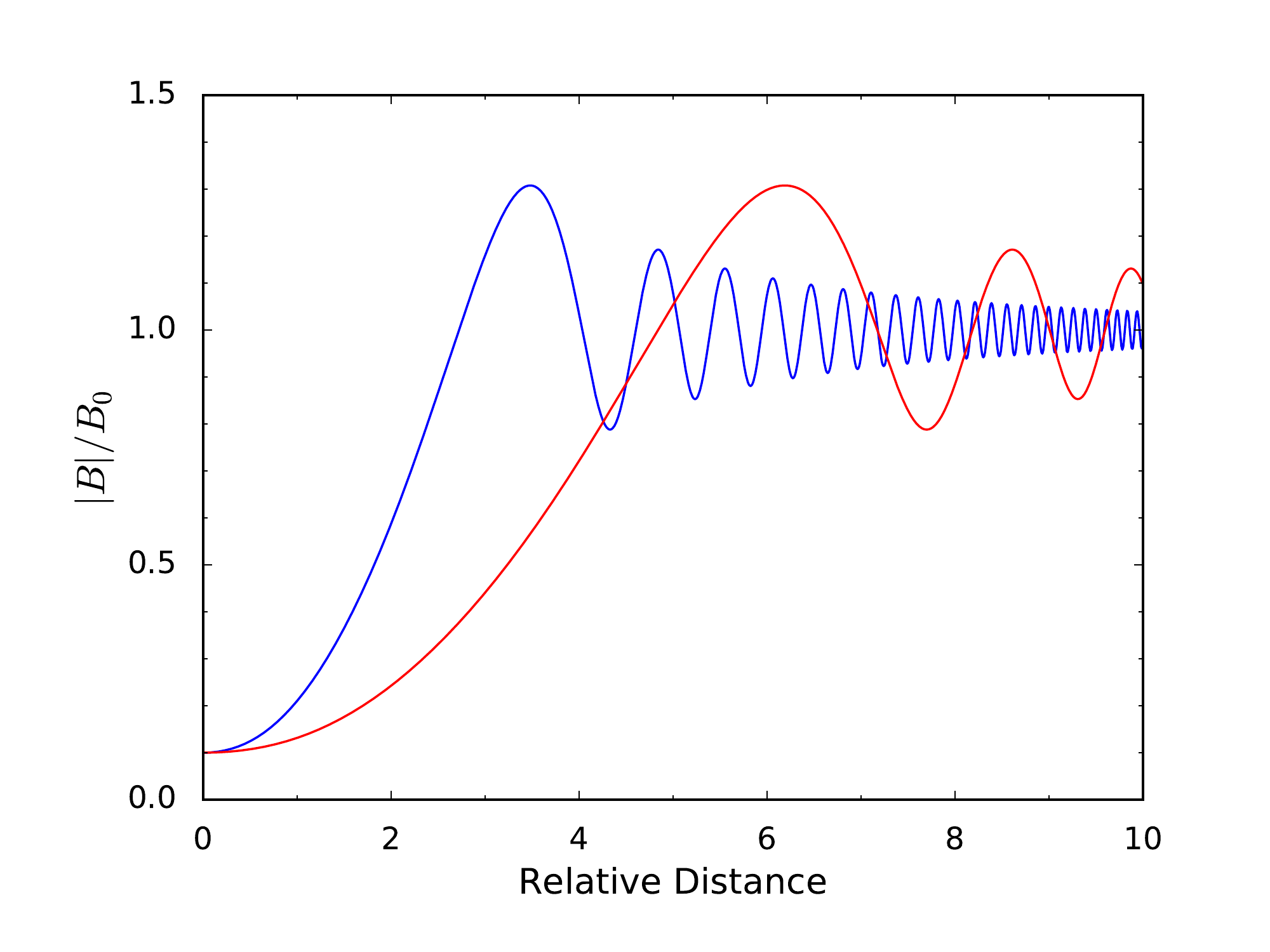} 
\end{tabular}
\caption{Profile of 
the horizontal magnetic field 
$|B|$ at two different times,
according to the self-similar solution given by \Eq~(\ref{fre2}). The peaks are
moving from left to right.}
\label{fresnel2}
\end{figure}

The analytical solution for the Hall evolution caused by the jump
is given by
\begin{equation}\label{fre2}
	B(z,t)=B_1+ \frac{(8/\pi)^{1/2}}{1-i}\left[\mathcal{C}(u)-i\mathcal{S}(u)\right](B_0-B_1),
\end{equation}
where $u = (z-z_0)/\sqrt{4D_H (t-t_0)}$ and $t_0$ is the initial time at which the jump was 
created. Snapshots of this solution at $z>z_0$ are shown in Figure~\ref{fresnel2}.
This solution is similar to Equation (\ref{fre1}), but with different boundary conditions.
The peaks of the oscillating profile
are moving from left to right 
and their widths increase
with time as $t^{1/2}$. In the idealized problem, where the infinitely 
sharp jump is created instantaneously, the peaks start out infinitesimally narrow.
More realistically, the jump is implanted at the end of a plastic failure in a finite time 
$\delta t\sim \alpha^{-1}$. This timescale determines the characteristic peak width at 
the beginning of the evolution. At times $t-t_0\gg\delta t$ the evolution becomes 
self-similar and accurately described by \Eq~(\ref{fre2}).

A key feature is that {\it the peaks of the launched Hall waves significantly exceed 
the background field $B_0$}. 
The highest peak is $1.3B_0$. This implies that the Hall waves 
are capable of breaking the crust and inducing new plastic flows, leading to an avalanche
of plastic failures.
\footnote{While we believe that this effect will also be present in multi-dimensional configurations, geometry and non-linearities present in multi-dimension can change it quantitatively}

The avalanche development can be demonstrated by the following numerical experiment.
Consider a uniform crust with electron density $n_e = 10^{35}$~cm$^{-3}$ and 
$B_z=3\times 10^{14}$~G; this gives the Hall diffusion coefficient 
$D_H \approx 0.015$~cm$^2$~s$^{-1}$.
To isolate the effect of interest we turn off heat diffusion, so there will be no TPWs, 
and new plastic flows can only be induced by Hall waves. As an initial condition at $t=0$ 
we take a uniform field $|B_0|=2.7\times 10^{14}$~G. We set 
$\Bcr= 3\times 10^{14}$~G everywhere except a small region $|z-z_0|<2$~m. 
In this region, we trigger the plastic flow by setting $\Bcr^{\rm new}=0.1\Bcr$. 
This setup is designed to produce an initially localized plastic flow, which 
reduces the field in the small region and launches Hall waves. Our experiment follows the evolution
by solving the Hall equation (\ref{eq:evol})  and simulating any new plastic flows, which must be
triggered wherever $|B|$ exceeds $\Bcr$.

\begin{figure}[t]
\vspace*{-0.3cm}
\hspace*{-0.5cm}
\begin{tabular}{c}
\includegraphics[width=0.53\textwidth]{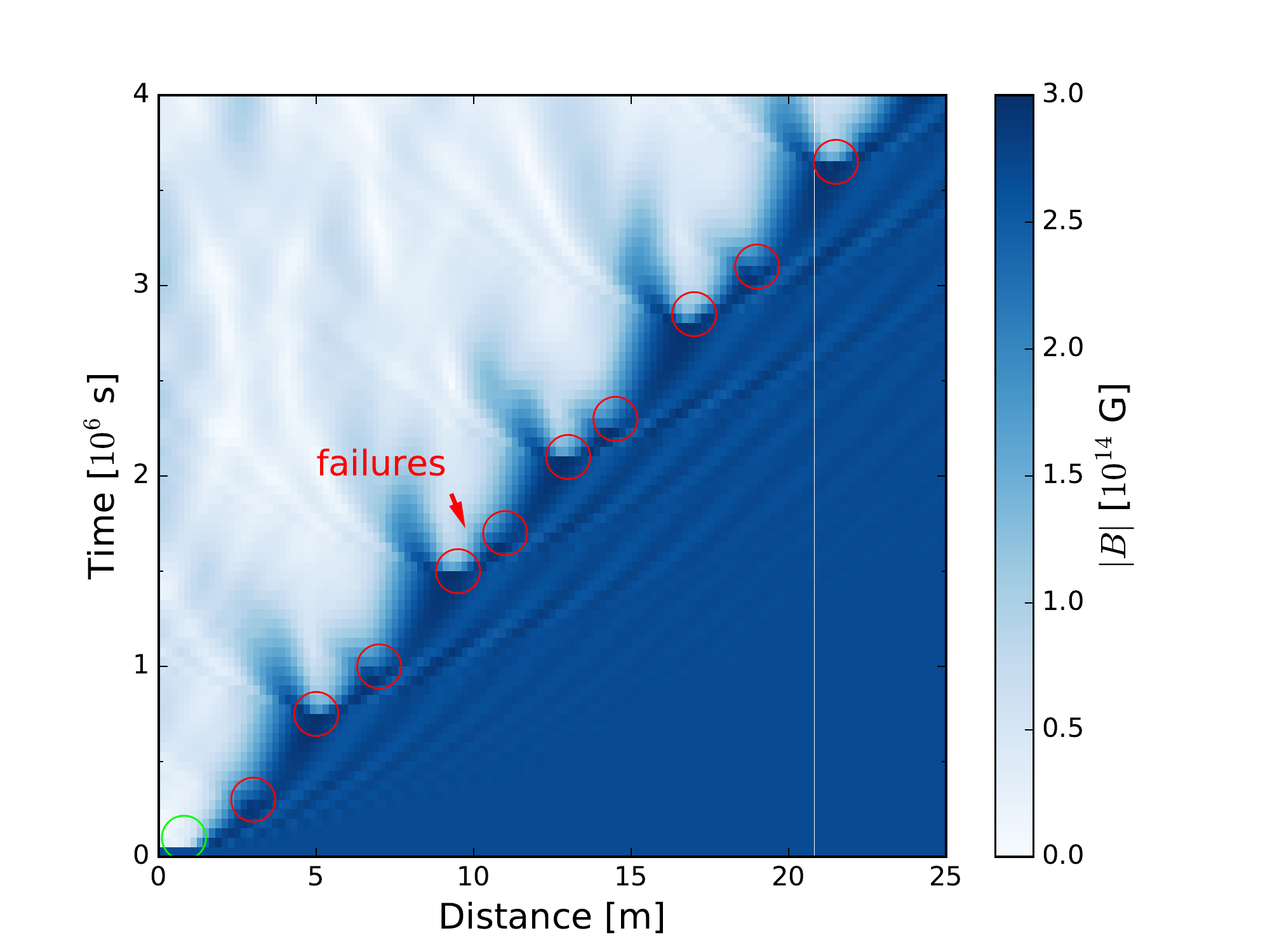} 
\end{tabular}
\caption{
Numerical simulation demonstrating the Hall-mediated failure of the crust
(see text). A seed plastic failure
is initiated at $t=0$ and $z<2$~m; it is indicated by the green circle on the spacetime 
diagram. Magnetic field $|B|$ is quickly reduced in the failed region, so it becomes 
white (weak field) according to the color code indicated next to the diagram. 
The failed region is expanding due to the short Hall waves launched
by the plastic flows into the intact region, 
where the waves trigger new plastic flows.
}
\label{failure}
\end{figure}

The result is convenient to view on the spacetime digram (Figure~\ref{failure}) that shows the 
evolution of $|B|$ in the region $z>z_0$. As expected, the peaks of launched Hall waves 
break the crust, creating new plastic flows. These flows generate new jumps in $B$, which 
create new Hall waves etc., expanding the region where the crust has failed. 
The average speed of the failed region expansion is determined by the slope 
of the boundary $z_{\rm front}\approx vt$ observed on the diagram. The magnetic field 
$|B|$ has been reduced (and magnetic energy has been dissipated) in the region 
$z<z_{\rm front}$. In contrast, in the region $z>z_{\rm front}$, the field is only perturbed by the faster and weaker 
Hall waves, which are seen as the propagating oscillations. 
The front speed is $v\approx 5.5\times 10^{-4}$~cm/s, about half of 
$(\alpha D_H)^{1/2}=1.2\times 10^{-3}$~cm/s --- the characteristic speed of
Hall waves launched by plastic flows. As a rough estimate one can use
\beq
\label{eq:vH}
  v\sim (\alpha D_H)^{1/2}.
\eeq
In essence, we observe a {\it Hall-mediated mode of crustal failure} (HMF). 
It is much slower than the heat-mediated TPW, however it can operate where 
TPWs do not propagate.

The HMF expansion occurs with 
no clean separation of the timescales of Hall and 
plastic evolution --- the frequencies of excited Hall waves are comparable to $\alpha^{-1}$.
Therefore, the details of failure propagation are complicated. In particular, we 
observe a curious limit-cycle behavior:
a fast reduction of $|B|$ on the timescale $\alpha^{-1}$ is followed by a failure
event with a slower reduction of $|B|$ in a smaller region; then it takes a long 
time to trigger a new failure which again turns out fast and strong, closing the cycle.

The cycle is explained as follows. A fast failure launches a Hall wave with
a narrow peak, $\delta z\sim (D_H/\alpha)^{1/2}$, and therefore the next plastic 
flow triggered by this peak occurs promptly and in a narrow region.\footnote{Plastic 
     flow is initiated where the growing peak touches $\Bcr$, and the width of the induced 
     failure is controlled by the curvature of $|B(z)|$ at the peak.
     In a complete model, with heat diffusion,
     a local TPW is produced, which quickly extinguishes as it propagates
     away from the peak. The model with switched off heat diffusion does not generate TPWs; 
     instead, it generates a cascade of very thin plastic flows (limited by the grid resolution of
     $2$~cm), which merge with time. Thus, the behavior on the smallest scales cannot be 
     resolved in the simulation presented in Figure~4. However, the small-scale details weakly 
     affect the behavior
     of the front on scales well above the grid scale --- it turns out similar to 
     a more complete simulation with included heat diffusion.
     }    
The relaxation of $|B|$ to $\Bcr^{\rm new}=0.1\Bcr$ in the narrow plastic region
is hindered by the Hall-wave transport of magnetic energy
--- the locally dissipated magnetic energy is replenished 
by the Poynting flux 
\begin{eqnarray}
F_p&=&\frac{1}{8\pi} i D_H(B^*\partial_z B-B\partial_z B^*) \nonumber\\
 &=&\frac{1}{4\pi}D_H(B_y\partial_z B_x-B_x\partial_z B_y)
\end{eqnarray}
into the plastic region, which resists the development of a localized
sharp drop in $|B|$.
The delay in the drop of $|B|$ causes a delay in the launching of a new super-critical 
Hall wave into the intact region ahead of $z_{\rm front}$.
As a result the next failure event at $z>z_{\rm front}$ is 
delayed. When it finally occurs the wave peak is broad and triggers a plastic flow in a 
relatively broad region.  The relaxation $|B|\rightarrow \Bcr^{\rm new}$ in the broad 
region is not hindered by the Poynting flux and occurs promptly, on the timescale of 
$\sim \alpha^{-1}$.

The limit cycle can be clearly seen in
Figure~\ref{failure} as the repeating appearance of ``fingers'' and fast failures.
A ``finger'' pattern (e.g. one near distance $4$~m and time $10^6$~s) contains narrow failures separated by intact regions. 
Magnetic energy is only dissipated at narrow failure sites,
and the Poynting flux from intact regions 
replenishes the energy there.
It takes longer for $|B|$ to drop below $B_{\rm cr}^{\rm new}$ in the ``finger'' compared to fast failures to the right.
After a fast failure in a broad region, new narrow failures are reproduced. Therefore the ``finger'' appears again.

We also performed simulations similar to that shown in Figure~\ref{failure} that included 
heat diffusion and ohmic dissipation with a realistic $\eta\sim D_H/20$. 
We observed a similar propagation of the failure front, but with gradually damped peaks
of the Hall waves. We also varied $|B_0|/\Bcr$ and found that when this ratio is closer to 
unity, 
the TPW becomes the dominant mode of failure propagation. It propagates
much faster, with the velocity $v_{\rm TPW}\sim (\alpha\chi)^{1/2}$ 
well above
the velocity of the Hall-mediated failure $v_{\rm HMF} \sim(\alpha D_H)^{1/2}$.
Both failure modes will be seen to occur in the magnetar crust simulated in Section~\ref{sim}.

\section{Twisted external field}\label{twist}

The crustal motions  
must twist the external magnetosphere attached to the crust. Such external twists are 
observed in persistent magnetars through their hard X-ray emission 
\citep{2013ApJ...762...13B, 2014ApJ...786L...1H}.
In transient magnetars, evidence for magnetospheric twists is 
provided by shrinking hot spots on the stellar surface \citep{2009ApJ...703.1044B}.

The external twist implies a non-zero horizontal magnetic field at the stellar surface 
$B_s\neq 0$. This changes the stress balance inside the crust, which now reads
\beq
   \frac{\tilde{B} B_z}{4\pi}=-\mu\partial_z \xi_{\rm ela}, \qquad \tilde{B}=B-B_s.
\eeq
It leads to a modified version of \Eq~(\ref{eq:evol}) for the magnetic field evolution, 
\beq\label{fulleqn}
\left(1+\frac{\mu_B}{\mu}\right)\dot{B}= -i\partial_z\left(D\partial_z B \right)+\frac{B_z^2}{4\pi\mu}\dot{B_s}+B_z\partial_z\dot{\xi}_{\rm pl}.
\eeq
Plastic flows are triggered where $\tilde{B}$ exceeds $\Bcr$, and $\tilde{B}$ should 
replace $B$ in the equation of plastic flow dynamics. Therefore, 
\Eq~(\ref{plasticprescription}) is replaced by
\beq
B_z\partial_z \dot{\xi}_{\rm pl}=-\alpha \tilde{B}\left(1-\frac{B_{\rm cr}^{\rm new}}{|\tilde{B}|}\right)\Theta\left(|\tilde{B}|-B_{\rm cr}^{\rm new}\right).
\label{plasticprescription1}
\eeq

To close the set of equations describing the system, one must specify the evolution of $B_s$. 
It is controlled by two factors: (1) $B_s$ is pumped by the crustal motions.
(2) The external twist has a finite lifetime, because it requires a magnetospheric current
${\mathbf j}=(c/4\pi)\nabla\times{\mathbf B}\neq 0$. For instance, in the axisymmetric geometry $B_s\neq 0$ would be a toroidal field 
which requires a poloidal 
magnetospheric current \citep{2002ApJ...574..332T}. The current is sustained through 
$e^\pm$ discharge with a threshold voltage that regulates the damping time of the 
external twist to $\sim 1$~yr \citep{2007ApJ...657..967B}.

The pumping of $B_s$ by crustal motions can be implemented 
in our one-dimensional model as shown in Figure~5.
We consider a magnetosphere with a constant vertical magnetic field $B_z$ and a 
constant horizontal field $B_s$
that serves as a proxy for the twist component in three dimensions.
The pumping of $B_s$ is caused by the motion of magnetic field lines at the 
crust surface. This motion can be related to the evolution of $B(z,t)$ inside the crust
by integrating \Eq~(\ref{vel}) from the bottom to the top of the crust,

\beq
\int\limits_{\rm bottom}^{\rm surface}\md z\,\left.\frac{\dot{B}}{B_z}=(v+\eta\partial_z B)\right|_{\rm bottom}^{\rm surface},
\eeq
where $v(z,t)=v_H+\dot{\xi}=v_x+iv_y$ is the velocity vector of the electron fluid. 
Neglecting the lattice deformation 
at the base of the crust, 
$\dot{\xi}_{\rm bottom}=0$, we get 
the surface velocity of the magnetic field lines,
\beq
v_s=\int\limits_{\rm bottom}^{\rm surface}\md z\,\frac{\dot{B}}{B_z}
+\left(v_H+\eta\partial_z B\right)_{\rm bottom}.
\label{vsurf}
\eeq
Here we kept the resistive term $\eta\partial_z B$ at the bottom (because the typical 
setup of our simulations has a strong current sheet at the bottom),
and neglected $\eta\partial_z B$ at the surface, as its effect on the evolution of $B_s$ 
will be small compared to ohmic dissipation in the magnetosphere.
The magnetic field at the surface follows
the motion of the electron fluid with
velocity $v_s$.

\begin{figure}[t]
\vspace*{-0.4cm}
\hspace*{-0.5cm}
\begin{tabular}{c}
\includegraphics[width=0.53\textwidth]{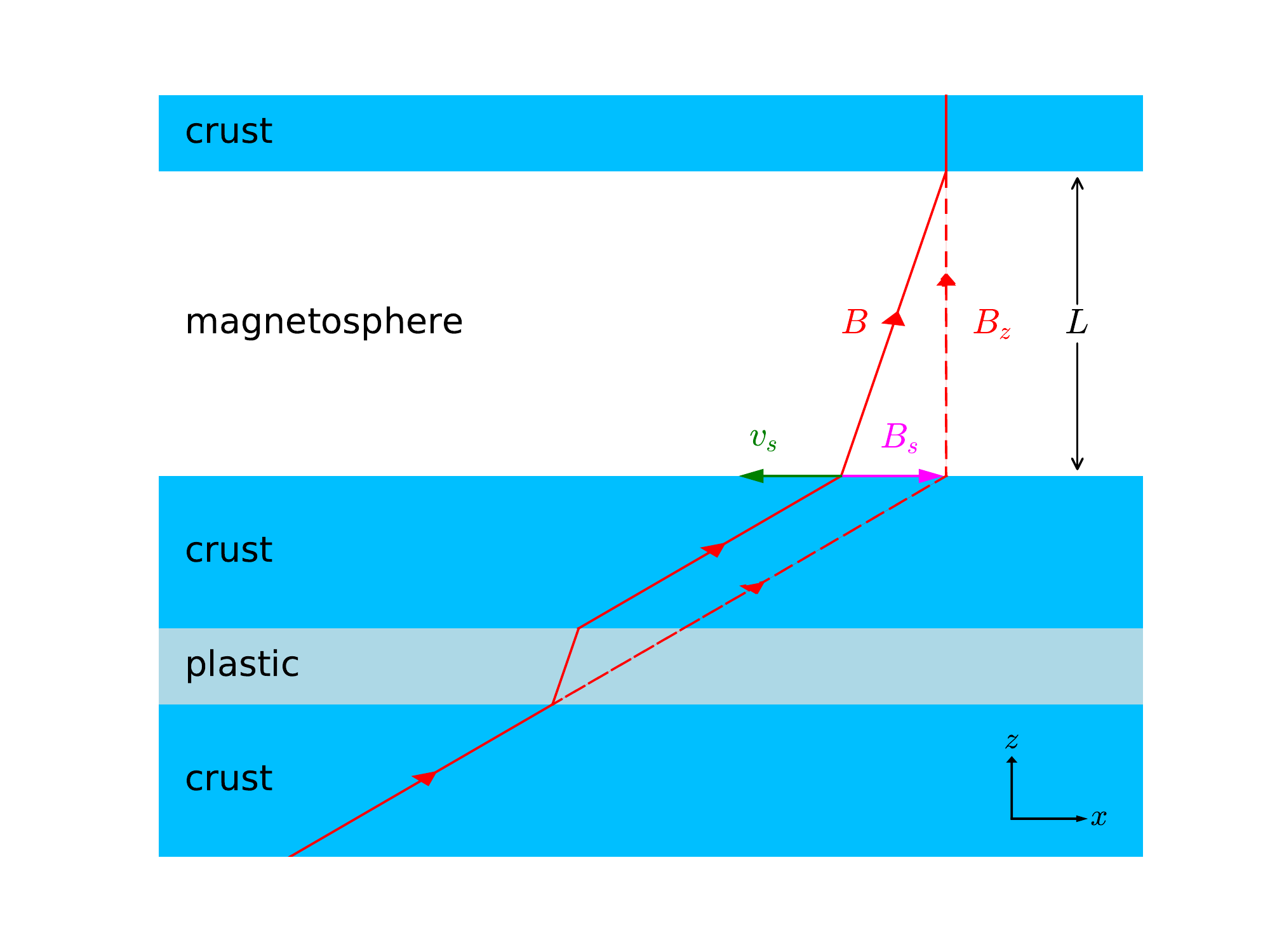} 
\end{tabular}
\caption{Illustration for the magnetic field lines in the crust (blue region) and in the 
magnetosphere (white region). 
A closed magnetic field line is anchored between two footpoints in the crust. The plot is projected in the direction of magnetic field line.
Plastic flow 
occurs
in the light blue region and 
results in a crustal surface motion
with velocity $v_s$. This motion shears the magnetosphere 
and creates $B_s$.
Red dashed and solid lines are the magnetic field line before and after 
the plastic deformation. 
}
\label{fig1}
\end{figure}

The rate of pumping $B_s$ is proportional to
$v_s$, and the evolution equation for $B_s$ may be written in the form,
\beq\label{bs}
\dot{B}_s=-B_z\frac{v_s}{L}-\frac{B_s}{\tau_{\rm damp}}.
\eeq
Here $L$ represents the length of the magnetospheric field lines (Figure~\ref{fig1})
and $\tau_{\rm damp}$
is the damping timescale. In a complete model, the value of $\tau_{\rm damp}$ would
depend on the voltage of $e^\pm$ discharge in the magnetosphere and the geometry 
of the twisted bundle of field lines \citep{2009ApJ...703.1044B}.  In our simplified model we 
fix $\tau_{\rm damp}=1$~yr.

The Hall evolution at the bottom boundary is slow and not capable of pumping 
$B_s$ against the twist damping in the magnetosphere. 
However, significant external twists can 
be created as a result of the large $\dot{B}$ in the regions of plastic 
failures \citep{2014ApJ...794L..24B}. 

\section{Global Simulation}\label{sim}

\subsection{Setup}

We now collect all the ingredients described in the previous sections into a 
global simulation of the magnetic field evolution in a magnetar crust over $10$~kyr.
The crust will now have a realistic density profile $\rho(z)$. Its temperature profile $T(z)$
is calculated self-consistently by evolving the time-dependent equation for heat transfer,
\beq\label{heat}
 C_V\dot{T}=\partial_z\left(\kappa \partial_z T\right)+q_{\rm pl}+q_{\rm ohm}-q_{\nu}.
 \eeq
It takes into account the plastic heating (\Eq~\ref{qpl}), the ohmic heating 
$q_{\rm ohm}=(\eta/4\pi)|\partial_zB|^2$,
and the energy losses due to neutrino emission $q_\nu$.

The coupled evolution of $B(z,t)$ and $T(z,t)$ in the crust, and $B_s(t)$ at the surface, 
is followed by solving Equations~(\ref{fulleqn}), (\ref{heat}), (\ref{vsurf}), and (\ref{bs}); 
\Eq~(\ref{plasticprescription1}) is used where plastic flows occur.

While our one-dimensional model can only approximate the behaviour of a real magnetar, it allows one to use
a realistic vertical profile for all of the important physical parameters of the crust.
We use the BSK20 model provided by \citet{2013A&A...560A..48P} to compute the density $\rho$, electron number density $n_e$, nuclear charge $Z$ and mass $A$ as a function of depth $z$. 
The mass  of the neutron star is chosen to be $1.4M_{\odot}$, for which the model predicts the radius of  11.7~km.
The shear modulus $\mu$ is calculated using the fitting formula provided by \citet{2005ApJ...634L.153P} and \citet{2007MNRAS.375..261S} for low and high densities. 
To accelerate the computations, in most of our runs the ohmic diffusivity  is set to $\eta = |D_H|/20$, the value which is characteristic for the inner crust. We have separately checked that the 
details of  
ohmic dissipation do not affect our results, as most of the energy is dissipated in
the plastic flow regions.  Our fiducial value for the vertical magnetic field is $B_z = 3\times 10^{14}$~G, which is a typical poloidal 
field of magnetars  
inferred from their spindown rates.

We choose the upper boundary of our simulation domain 
at $z=z_b$ where $\rho_b\equiv\rho(z_b)=10^{9}$~g/cm$^3$. Our results are not sensitive to this choice so long as 
(1)
the crustal shear modulus at the boundary is sufficiently weak, $\mu(\rho_b)\ll B_z^2/4\pi$, and (2) the 
timescale of heat conduction from the boundary to the stellar surface $t_c(\rho_b)$ is much 
shorter than the typical conduction time across the crust, which is comparable to one year. 
The choice of $\rho_b=10^9$~g/cm$^3$ gives $4\pi\mu(\rho_b)/B_z^2\sim 10^{-4}$
and $t_c(\rho_b)\sim 10^6$~s for typical magnetar temperatures. Our computational box
includes the entire crust at $\rho>\rho_b$, which has a
thickness of about 1~km. We use $30,000$ evenly spaced grid points;
this gives enough resolution for capturing small-scale Hall waves.

We employ the
Crank-Nicolson scheme \citep{press2007numerical}
to solve both the Hall wave propagation and the thermal evolution.
In our fiducial run, we 
keep a constant horizontal magnetic field $B_{\rm core}=6\times 10^{15}$~G at the 
lower crust boundary $z_{\rm core}$. The horizontal field at the upper boundary $B_s$
evolves dynamically according to \Eq~(\ref{bs}), which provides 
a time-dependent boundary condition at $\rho_b$. The initial condition for the horizontal field
$B$ is chosen to be
\beq
B(z)=B_{\rm core}\exp\left[-\frac{(z-z_{\rm core})^2}{l^2} \right],
\qquad z_b<z<z_{\rm core},
\eeq
with $l=10$~m. 
Thus, initially the crust has practically no horizontal field, and the presence 
of a strong horizontal field at the boundary launches Hall waves into the crust as 
described in Section~\ref{generation}.
The exact value of $l\ll 100$~m
has no impact on the model's long-term behavior. We envisage that this type of initial 
configuration may result from a quick
rearrangement of the core magnetic field.

It is important for our purposes to accurately track the thermal evolution of the crust. 
We choose the initial surface temperature to be $T_{s0}=2\times 10^6$~K which is 
typical for transient magnetars in quiescence.
The initial temperature profile below the surface sustains the steady heat flux 
$F=-\kappa\partial_z T=\sigma_{\rm SB} T_{s0}^4$
conducted from the core. 
The corresponding temperature of the core and the lower crust is $\sim 3\times 10^8$~K.
Neutrino cooling is negligible at such temperatures, however it 
becomes important later when the crust is heated by the avalanches of Hall waves and
thermoplastic waves. 
For simplicity,
the core temperature is kept constant throughout the simulation.
This may be reasonable due to the high heat capacity of the core, and this also
assumes that the main phase of its intrinsic thermal evolution occurred at 
earlier times, see \citet{2016arXiv160509077B}.
The temperature profile of the heated crust is evolved by solving the time-dependent heat 
transfer equation as described in \citet{2015ApJ...815...25L}.
We use the code provided by \citet{1999A&A...351..787P} to calculate $C_V$ and $\kappa$ 
in the strong magnetic field. At high temperatures, the crust is 
efficiently cooled by neutrino emission.
Several processes contribute to the neutrino emissivity $q_\nu$; 
our simulations 
include the effects of annihilation of electron-positron pairs, plasmon decay, neutrino 
bremsstrahlung, and neutrino synchrotron emission. We use the formulae provided in 
\citet{2001PhR...354....1Y} to calculate $q_\nu$ from all these channels. 

The Hall wave propagation is typically slow compared to plastic instabilities and heat 
propagation. Therefore, the timestep in our simulations is adaptive and chosen to resolve the 
fastest processes when they happen --- the plastic flows, neutrino 
cooling, and twisting of the external magnetosphere. We require that
the local temperature change due to plastic heating or neutrino emission in one timestep
does not exceed $10^7$~K and the change in $B_s$ is smaller than $10^{9}$~G. 
To speed up the calculation, we track the thermal evolution only when there is an episode 
of plastic heating until the temperature profile has 
relaxed back to the steady heat flow from the core (i.e. when the temperature profile 
is everywhere close to the initial state, with deviations smaller than $10^7$~K). 
When the thermal evolution is turned off, the timestep is set 
at $5\times 10^4$~s. We have tested that
it is short enough to resolve the Hall wave propagation in the absence of plastic failures. 
We use $\alpha= 10^{-4}$~s$^{-1}$ and run our simulation to 10~kyr. The energy conservation 
is better than 1\% during the whole simulation for Hall wave evolution and 5\% for thermal evolution.

When no plastic failure is triggered, the crust is 
only
heated by ohmic dissipation.
Its effect on temperature is however small, $\Delta T<10^7$~K. When a plastic failure 
occurs, the local plastic heating greatly exceeds the ohmic heating. Therefore, we 
neglect the contribution of ohmic heating in the thermal evolution equation at all times. 
However, ohmic damping is taken into account in the 
Hall evolution equation for the magnetic field, where its effect is more significant.

As explained in Section~\ref{plastic}, an important element of our model is the 
dependence of the critical shear stress on temperature (the thermal softening of the crust). 
We use the expression given in \Eq~(\ref{shearstress}) as long as
there is no plastic failure. When the plastic flow is initiated, at failed sites
the maximal shear stress supported by the crust drops to $\sigcr^{\rm new}$,
10\% of its original value at zero temperature;
this corresponds to the reduction of $\Bcr$ by a factor of 10 (see Section~\ref{mechanical}).
The crustal lattice heals when $|B-B_s|B_z/4\pi$ approaches $\sigcr^{\rm new}$,
which occurs on the timescale of $3000/\alpha\sim 1$~year. At this point, the critical 
shear stress is increased back to the value given by \Eq~(\ref{shearstress}).
If the strong local heating melts the crust, then $\sigcr$ vanishes and the local $B$ 
must immediately relax to $B_s$, releasing magnetic energy. The dynamics of this fast 
process is not resolved in our simulations, instead we simply allow $B-B_s$ to be 
exponentially reduced on the timescale $\alpha^{-1}$ and convert the released 
magnetic energy to heat.

\begin{figure}[t]
\vspace*{-0.3cm}
\hspace*{-7mm}
\begin{tabular}{c}
\includegraphics[width=0.5\textwidth]{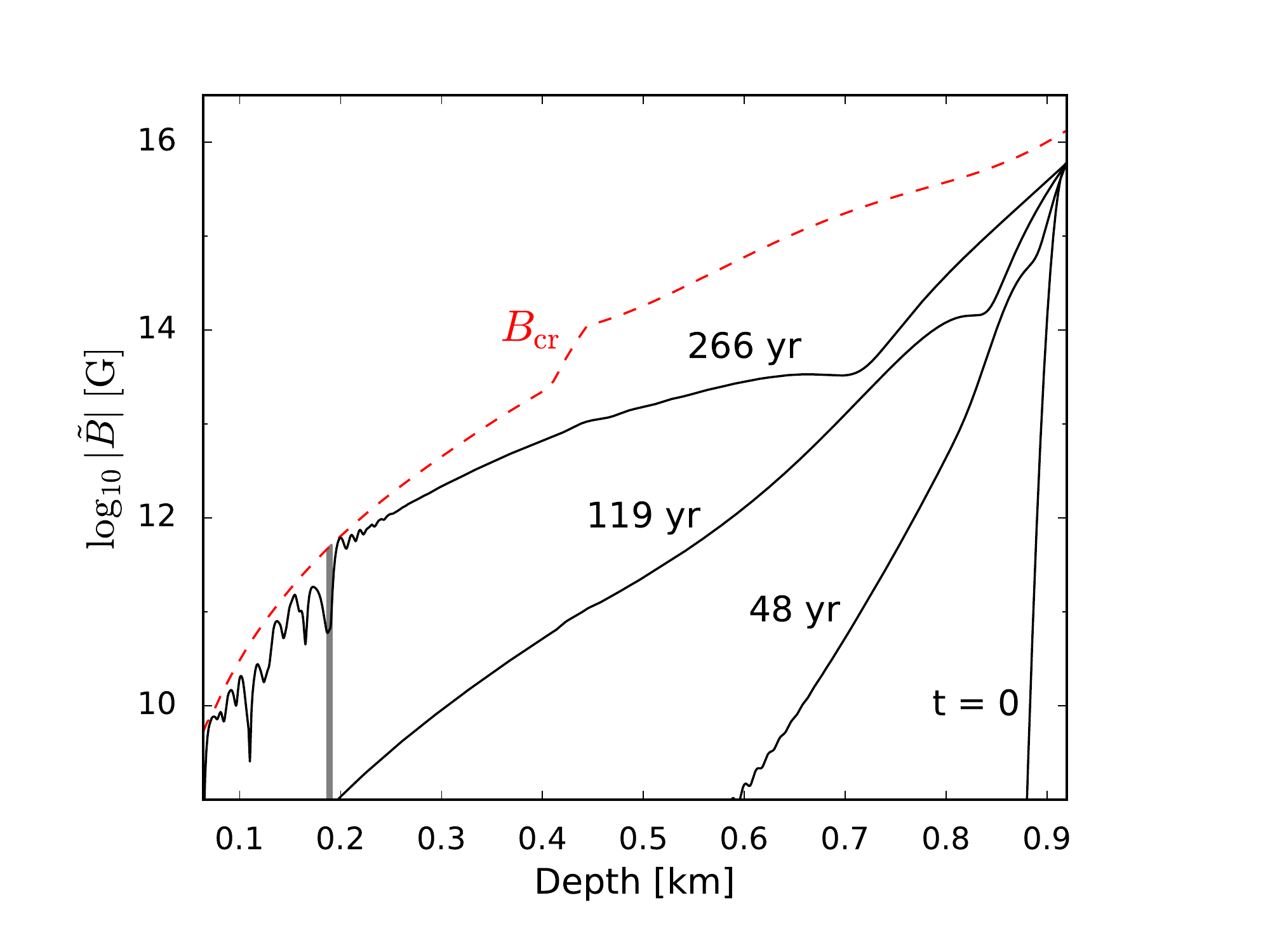} 
\end{tabular}
\caption{Snapshots of
the magnetic field evolution, showing how the horizontal magnetic field $B$ 
gradually fills the crust from its lower boundary at $z\approx 1$~km. 
The horizontal axis shows the depth of the crust. The neutron star surface is to the left and core-crust interface is on the right where Hall waves are launched.
Solid curves show $|\tilde{B}|=|B-B_s|$ vs. depth $z$.
$B_s\neq 0$ corresponds to the external magnetospheric twist; it is negligible 
at most times, except when the magnetosphere is quickly twisted by the plastic 
instabilities of the crust. Such motions are triggered when $\tilde{B}$ approaches 
$\Bcr$ (shown by the dashed red curve). This is seen to happen in the snapshot at 
$t=266$~yr; the plastically flowing region at $z\sim 0.2$~km is indicated by the vertical 
grey strip.
$\Bcr$ is calculated at the steady-state temperature profile that corresponds to the surface temperature of $2\times 10^{6}$~K.}
\label{fig2}
\end{figure}

\subsection{Results}

Figure~\ref{fig2} shows the initial Hall evolution, which gradually fills the crust with a 
horizontal magnetic field. 
The field eventually triggers plastic failures, which launch new Hall waves, and then 
the evolution continues in a chaotic manner, with repeating thermoplastic waves and 
Hall-mediated avalanches at various locations in the crust.

\begin{figure*}[htpb]
\vspace*{-0.3cm}
\hspace*{-0.6cm}
\begin{tabular}{c l}
\includegraphics[width=0.52\textwidth]{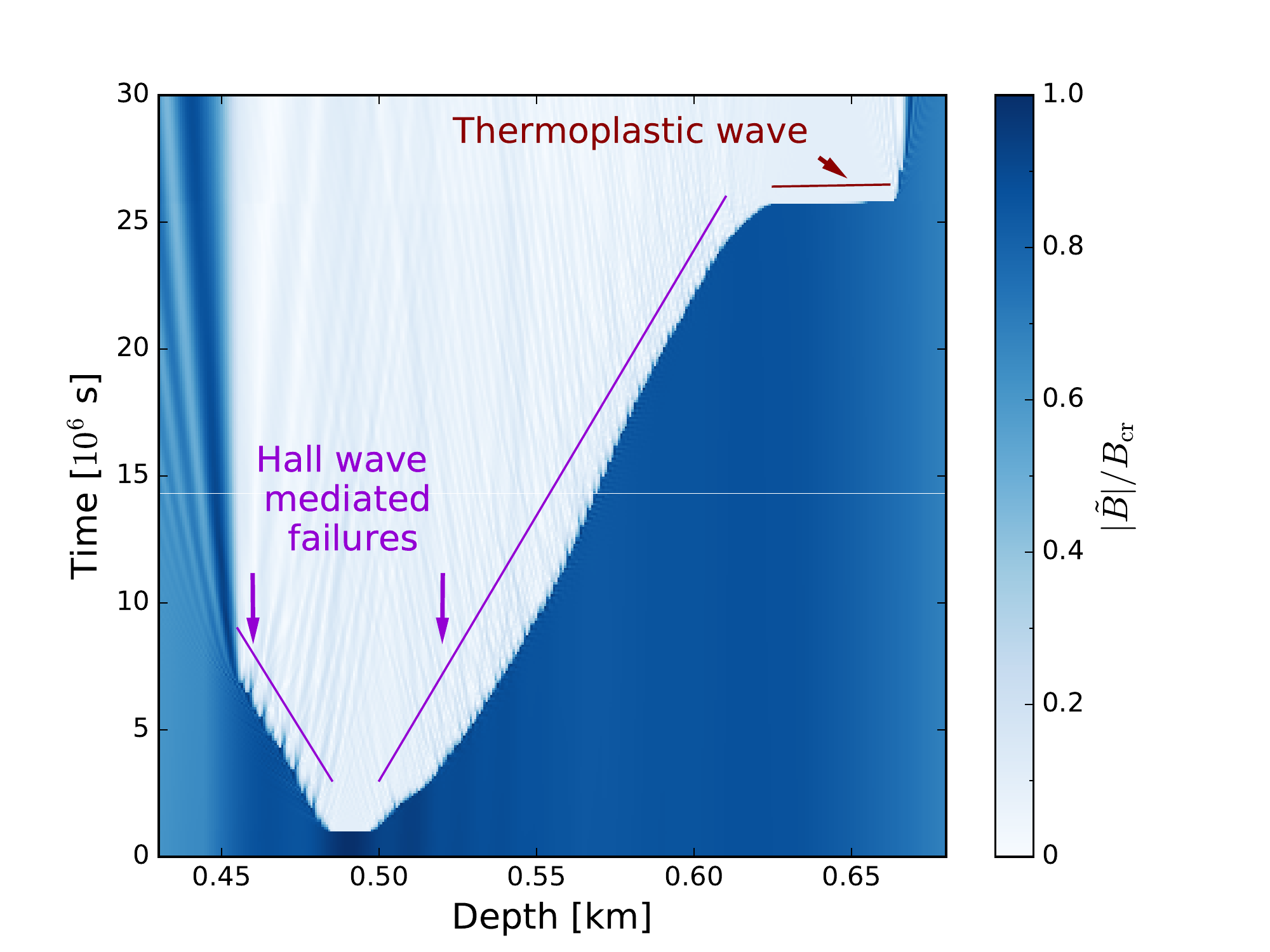}
&\includegraphics[width=0.52\textwidth]{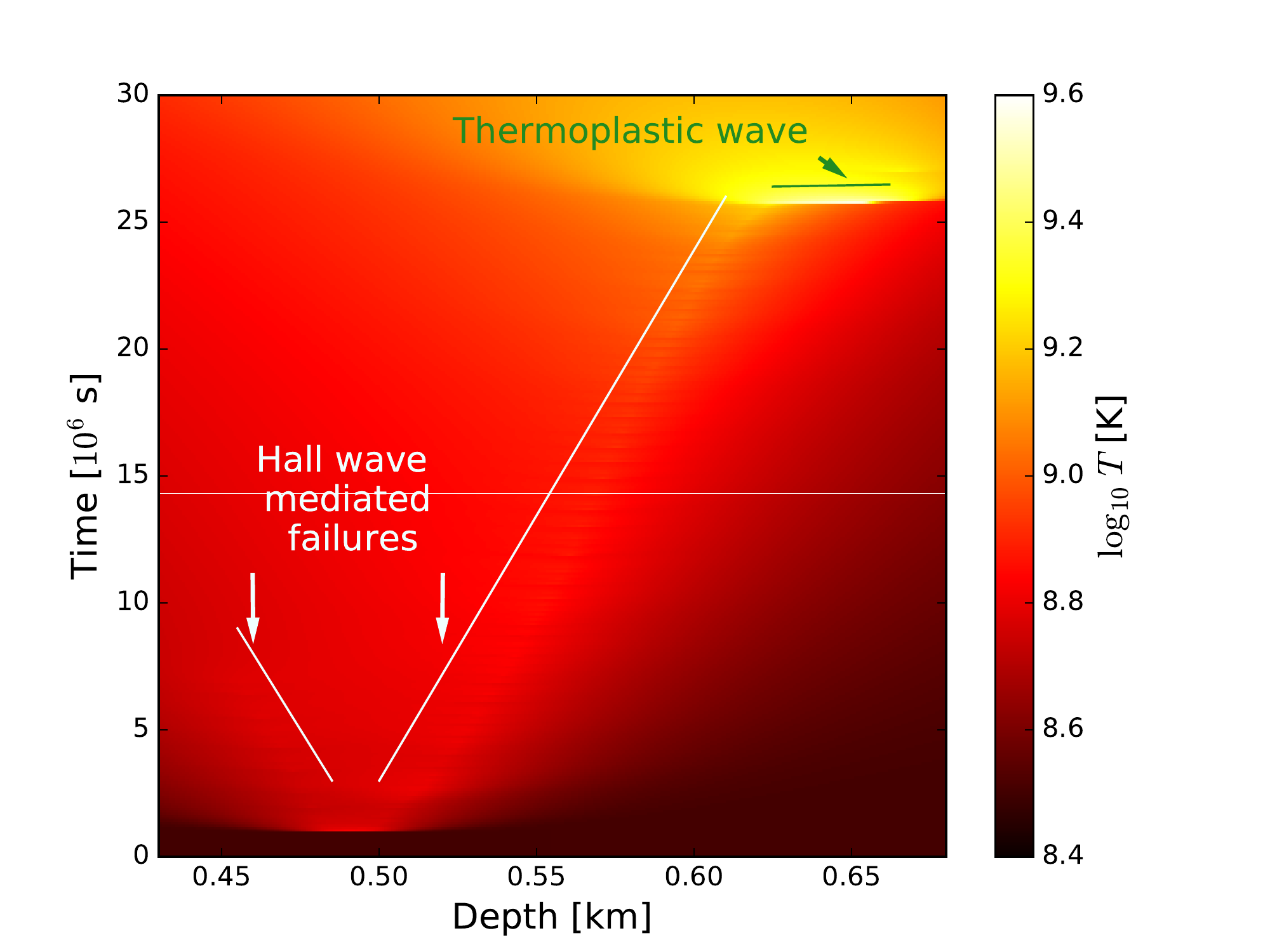} 

\end{tabular}
\caption{Spacetime diagram for failure generation and propagation in the crust. 
\textbf{Left panel:} evolution of $|\tilde{B}|/B_{\rm cr}$.
At the failure sites, the value of $|\tilde{B}|/B_{\rm cr}$ 
drops to 0.1 which 
corresponds to
the pale blue 
color.
The magnetic stresses in the neighborhood of the first failures (which are about 20~m thick)
are not able to launch large thermoplastic waves. Instead, propagation of the initial failures
is assisted by short Hall waves.
When the Hall-mediated avalanche reaches the depth of 0.6~km, a strong 
thermoplastic wave is launched, which propagates much faster and quickly reaches 
$z\approx 0.67$~km, where the wave extinguishes.
\textbf{Right panel}: 
Temperature evolution. Before the 
failure is triggered 
the temperature is kept 
near
the initial steady state. 
As the failure develops, plastic heating increases the 
temperature. The heating is particularly strong in the thermoplastic wave developing
at $z\approx 0.6$~km.
}
\label{sp2}
\end{figure*}

\begin{figure}[htpb]
\vspace*{-0.3cm}
\hspace*{-9mm}
\begin{tabular}{c}
\includegraphics[width=0.59\textwidth]{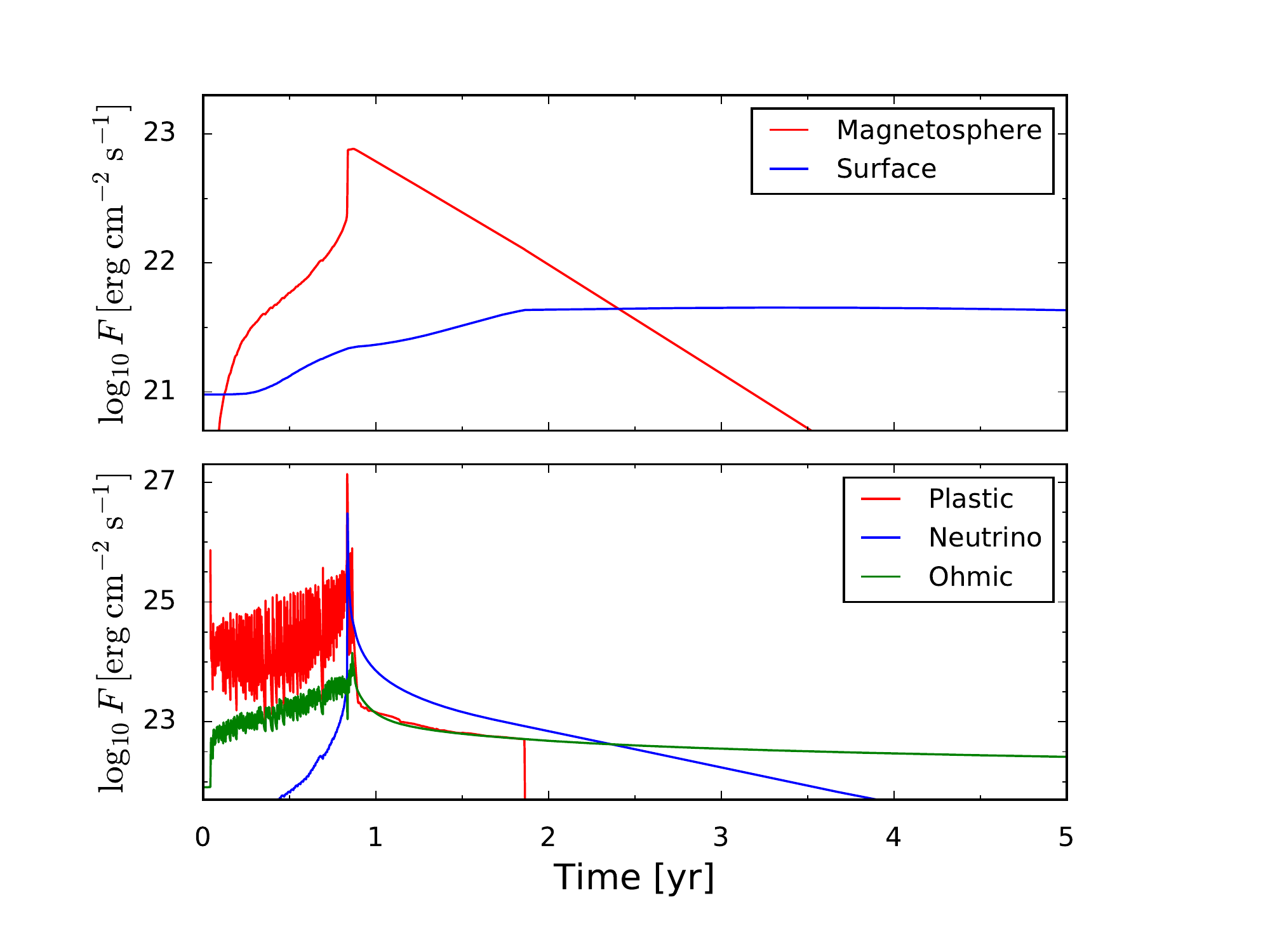} 
\end{tabular}
\vspace*{-5mm}
\caption{
\textbf{Upper panel:} radiation flux from the stellar surface (blue) and dissipation rate in the 
magnetosphere per unit area of the crust (red).
\textbf{Lower panel:} vertically integrated rates of plastic heating (red), ohmic heating 
(green), and neutrino cooling (blue), during and after the failure avalanche shown in 
Figure~\ref{sp2}.
}
\label{figz2}
\end{figure}

The spacetime diagram in Figure~\ref{sp2} shows failure development during a major 
Hall-mediated avalanche at $z\approx 0.5$~km, which propagates into the deeper crust 
and concludes with a strong thermoplastic wave at $z=0.6-0.67$~km.
The duration of the avalanche is about one year.
A significant magnetic and elastic energy is dissipated during this time through the friction 
in the plastic flow
(crustal ohmic heating 
makes a negligible contribution). The evolution of 
heating and neutrino cooling (integrated over depth $z$) is shown in the lower panel of 
Figure~\ref{figz2}. A fraction of the
produced heat is conducted to the stellar surface, increasing its luminosity. The evolution
of the surface radiation flux is shown in the upper panel of Figure~\ref{figz2}; it is rather 
smooth, because the characteristic timescale for heat conduction is comparable to one year.

The developing avalanche shears the stellar surface
and pumps magnetic energy into the magnetosphere. 
This energy is gradually dissipated through the continual $e^\pm$ discharge in the 
magnetosphere, producing radiation that is also shown in the upper panel of Figure~\ref{figz2}.
The magnetospheric activity rises slowly  during the Hall-mediated phase, and jumps upward 
when the strong thermoplastic wave occurs in the end of the avalanche.
This last event is quick and suddenly implants a significant twist into the magnetosphere.
After that, the magnetospheric emission decays 
resistively on the timescale of a year. 
Neutrino emission 
also peaks during the strong thermoplastic wave, because the temperature is highest
at this stage, and neutrino emission is extremely sensitive to temperature.

\begin{figure*}[t]
\vspace*{-0.3cm}
\hspace*{-0.6cm}
\begin{tabular}{c c}
\includegraphics[width=0.52\textwidth]{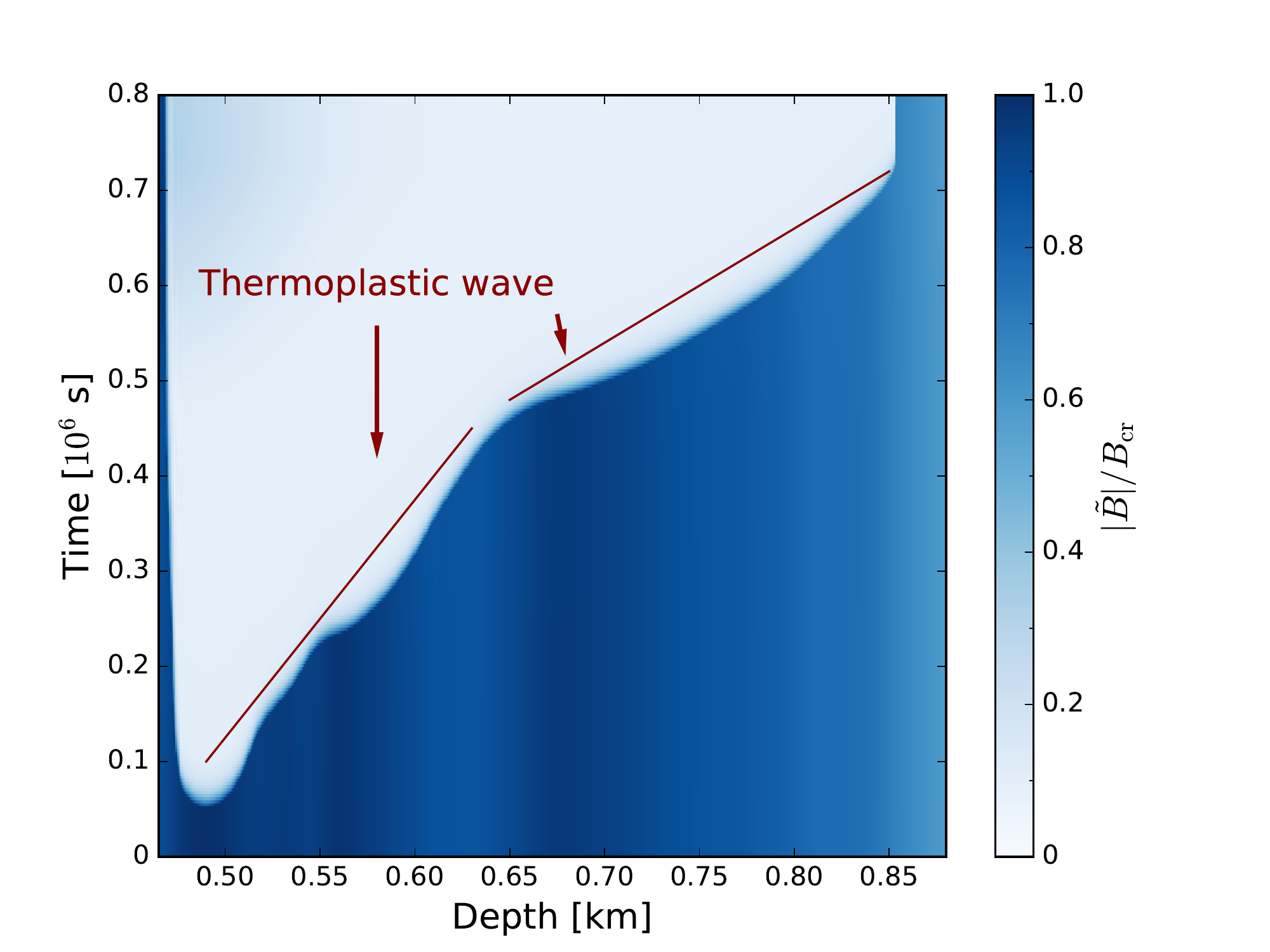} 
 &\includegraphics[width=0.52\textwidth]{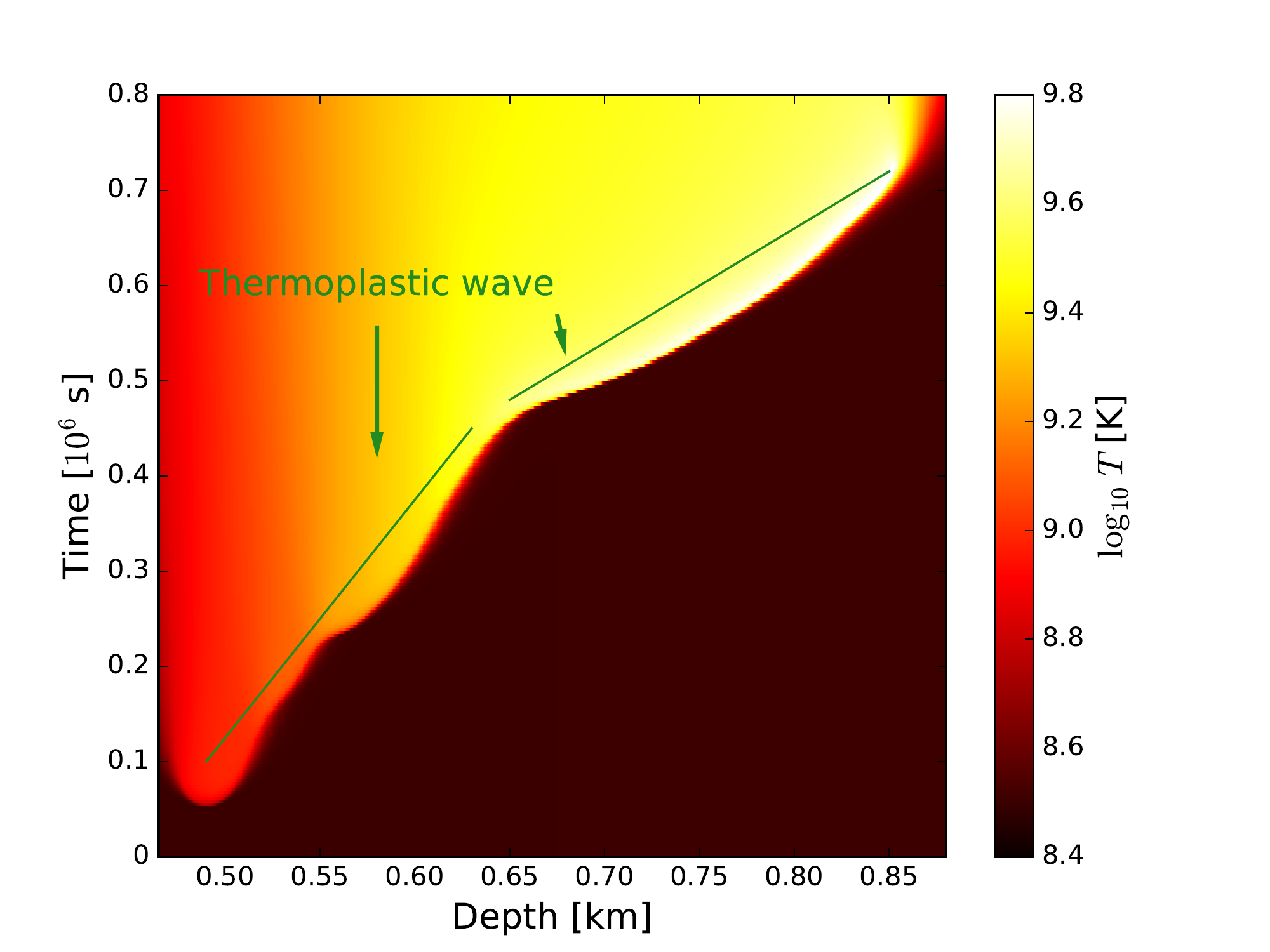} 
\end{tabular}
\caption{Spacetime diagram for a giant thermoplastic wave observed in the simulation at $t\approx 3.2$~kyr.
\textbf{Left panel:} evolution of $\tilde{B}/B_{\rm cr}$. 
Note the smaller scale on the time axis compared with Figure~\ref{sp2}; the
thermoplastic wave is much faster than 
the Hall-mediated avalanche. \textbf{Right panel:} temperature evolution. Compared to 
Figure~\ref{sp2}, 
the heating is stronger and occurs deeper in the crust.
There is also a strong temperature gradient across the wave front. 
}
\label{sp1}
\end{figure*}

\begin{figure}[t]
\vspace*{-0.3cm}
\hspace*{-9mm}
\begin{tabular}{c}
\includegraphics[width=0.57\textwidth]{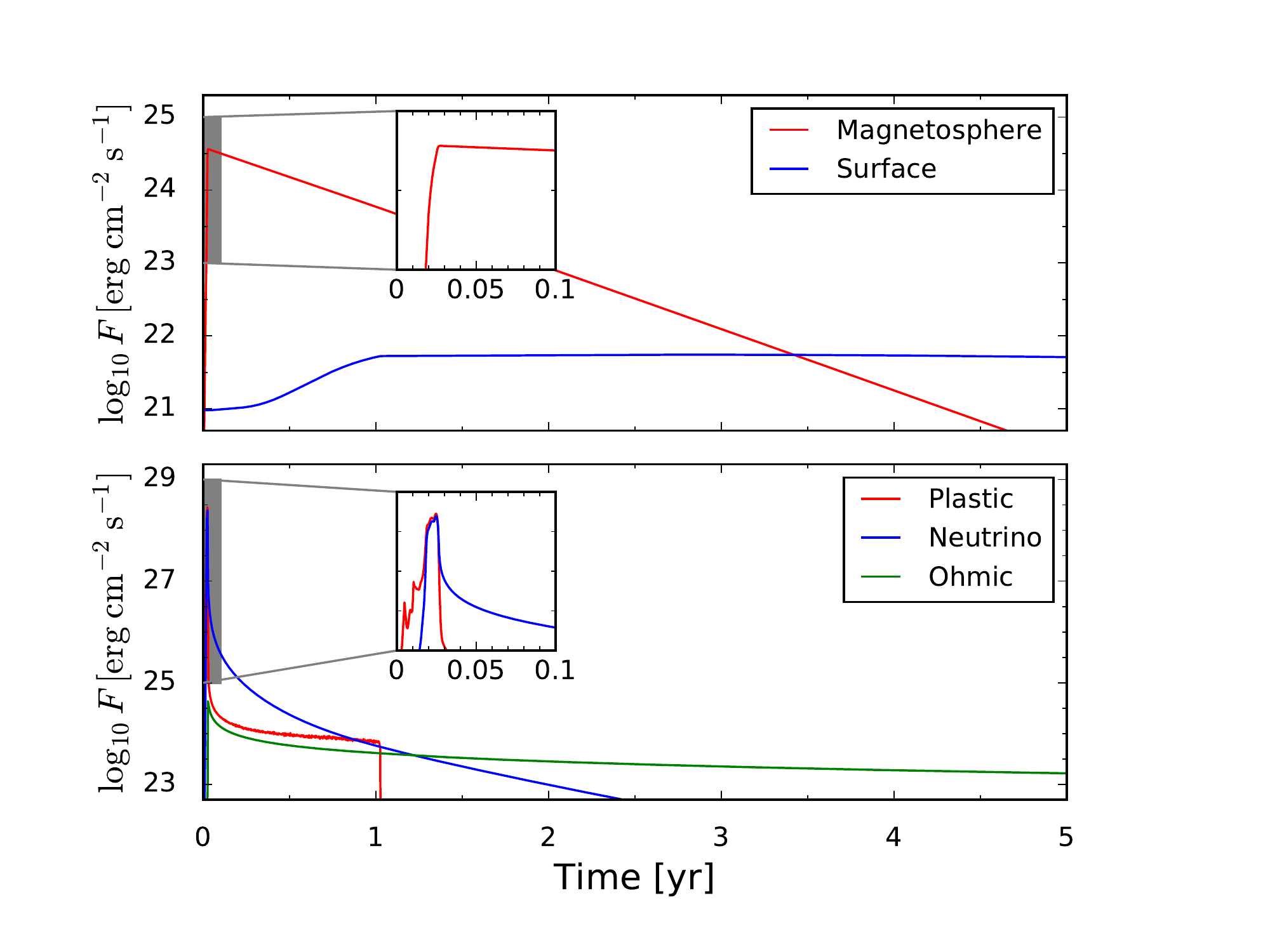} 
\end{tabular}
\vspace*{-0.4cm}
\caption{
\textbf{Upper panel:} radiation flux from the stellar surface (blue) and dissipation rate in the 
magnetosphere per unit area of the crust (red). The insets  zoom into the initial 
spike at the time of the thermoplastic wave.
\textbf{Lower panel:} vertically integrated rates of plastic heating (red), ohmic heating 
(green), and neutrino cooling (blue), during and after the thermoplastic wave shown in 
Figure~\ref{sp1}.
}
\label{figz1}
\end{figure}

Figure~\ref{sp1} shows another failure at $z\sim 0.5$~km at a later time during the evolution. 
This time, $\tilde{B}$ 
closely approached
$\Bcr$ in a broader range of depths,
and the failure immediately triggers a giant thermoplastic wave, which 
propagates from $z\approx 0.5$~km to 0.85~km. It travels much faster and 
a longer distance compared with the Hall-mediated avalanche in Figure~\ref{sp2}. 
The temperature is higher and there is a strong temperature gradient across the wave front,
which sustains its propagation.

Thermoplastic wave is a fast mode of failure propagation compared with the Hall-mediated
avalanche; in this example its duration is only 0.02 year ($7\times 10^5$~s). 
It produces fast and strong heating of the crust and twisting of the external 
magnetosphere (Figure~\ref{figz1}). However, the resulting radiation flux from the 
stellar surface is not much higher than in Figure~\ref{figz2}.
This is because heat is deposited deeper in the crust, and a larger fraction
of the heat is conducted into the core and lost to neutrinos.
This is in agreement with the behavior seen in \citet{2006MNRAS.371..477K,2014MNRAS.442.3484K}, see \citet{2016arXiv160509077B} for a discussion of the surface heating efficiency.

\begin{figure}[t]
\vspace*{-0.3cm}
\hspace*{-7mm}
\begin{tabular}{c}
\includegraphics[width=0.53\textwidth]{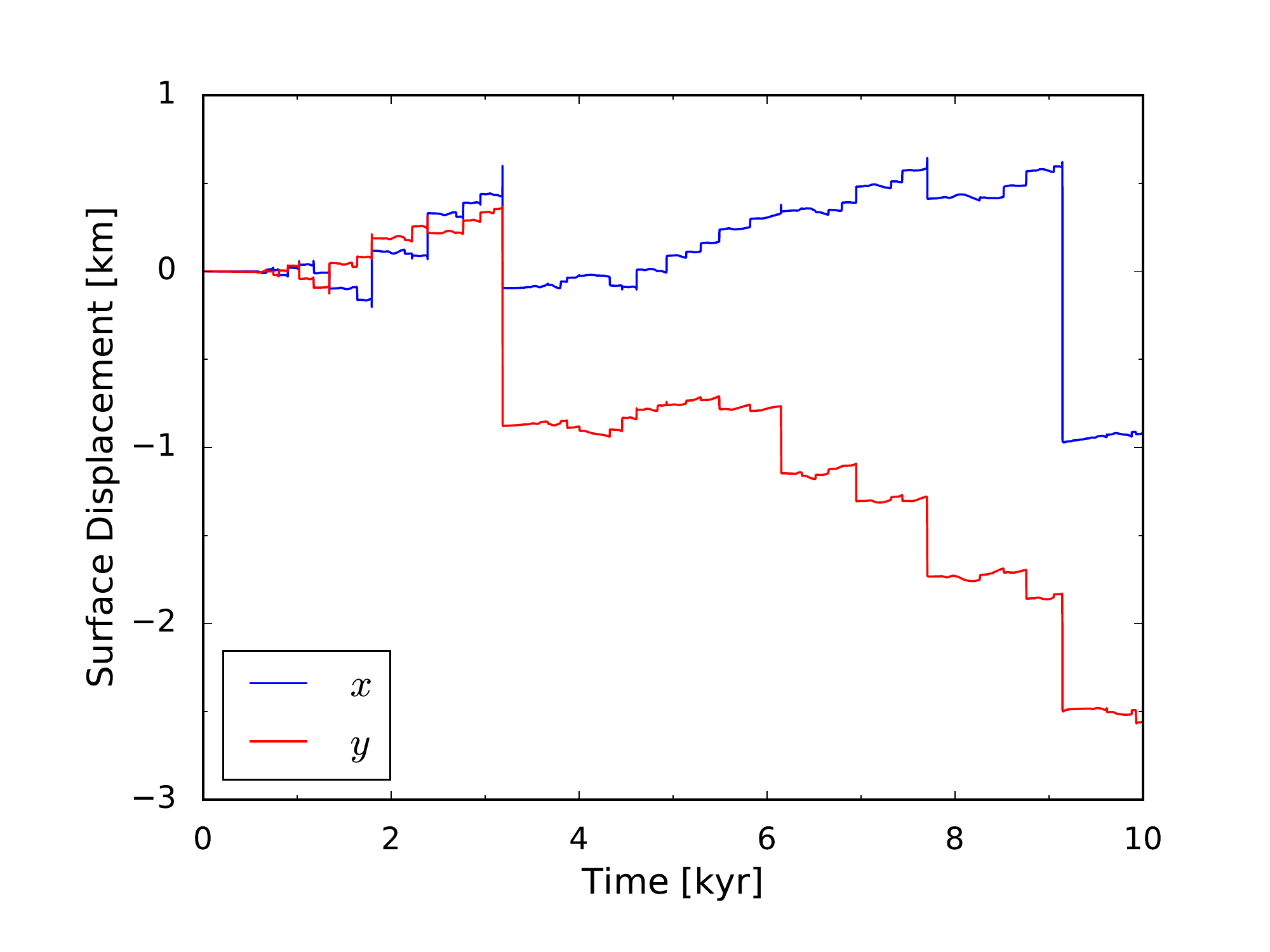} 
\end{tabular}
\vspace*{-0.4cm}
\caption{Evolution of surface displacement in the horizontal $x$ and $y$ directions
during the entire 10~kyr simulation.}
\label{dis}
\end{figure}

Figure~\ref{dis} shows the evolution of surface displacement in the $x$ and $y$ directions 
during our entire simulation. The displacement is zero at the beginning, before the Hall 
wave launched at the crust-core interface reaches the surface. Each failure event
causes the displacement to change abruptly. The large abrupt jump in the displacement  
near $3$~kyr is the result of the giant thermoplastic wave shown in
Figure~\ref{sp1}. The smaller jump 
near $7$~kyr corresponds to the event shown
in Figure~\ref{sp2}. There is another large jump of $1.5$~km caused by the thermoplastic 
wave at $t\approx 9$~kyr.

\begin{figure}[htpb]
\vspace*{-0.3cm}
\hspace*{-0.8cm}
\begin{tabular}{c}
\includegraphics[width=0.55\textwidth]{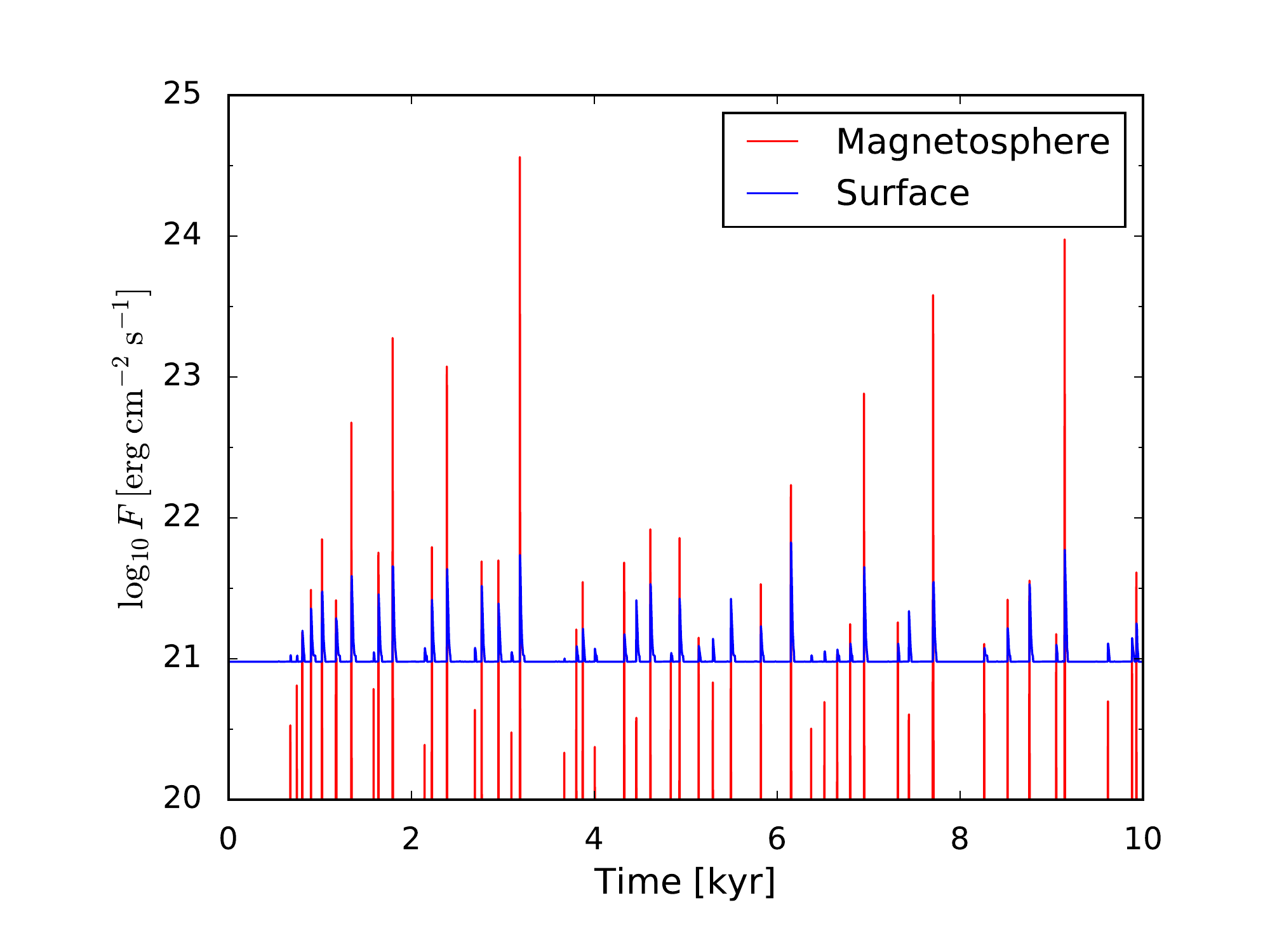} 
\end{tabular}
\vspace*{-0.3cm}
\caption{Evolution of 
the radiation flux from the stellar surface (blue) and dissipation rate in the 
magnetosphere per unit area of the crust (red) during the entire 10~kyr simulation.
}
\label{lum}
\end{figure}

Figure~\ref{lum} shows the evolution of observable radiation, from the surface and 
the magnetosphere, during the entire simulation. Both magnetospheric and surface 
emissions occur in sporadic spikes. 
There seems to be no obvious pattern for the spikes.
Each spike is an outburst caused by thermoplastic 
waves or Hall-mediated failures or a combination of both. The two large thermoplastic 
waves at $3$~kyr and $9$~kyr produce the strongest magnetospheric emission. 

\begin{figure}[htpb]
\vspace*{-0.3cm}
\hspace*{-8mm}
\begin{tabular}{c}
\includegraphics[width=0.55\textwidth]{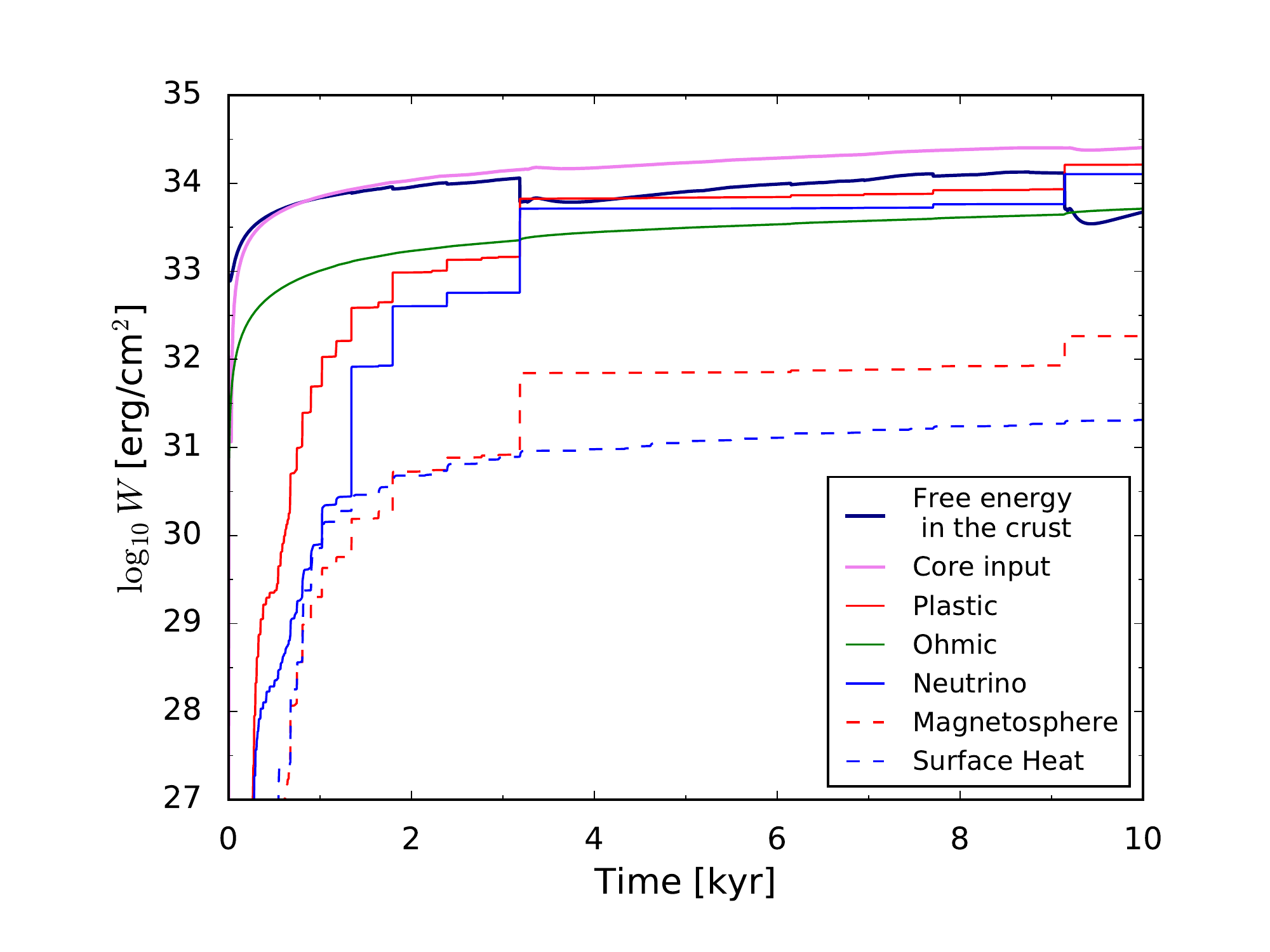} 
\end{tabular}
\vspace*{-0.3cm}
\caption{Evolution of the free energy stored in the crust and the contributions to its changes (see text).
}
\label{energy}
\end{figure}

Figure~\ref{energy} shows the evolution of 
the column density of ``free'' energy $W$ (density integrated over $z$) stored in 
the crust and available for dissipation; it includes the energy of the horizontal magnetic field 
and the elastic deformation energy.  The figure also shows the contributions to the changes in 
$W$ due to the Poynting flux from the core, plastic and ohmic dissipation, neutrino losses, 
the Poynting flux into the magnetosphere and heat flux radiated away at the surface above the persistent background.
Whenever there is an outburst, the free energy of the crust drops while the 
produced (time-integrated) heat, neutrino and radiative losses rapidly increase. 
Plastic and ohmic dissipation are the two main channels by which the crust is heated, with most of the heat lost to neutrinos or conducted into the core. While the plastic heating only happens sporadically, the ohmic heating takes place continuously at a much lower  rate and has a
negligible effect on the temperature change and emerging flux during the outburst. Only a small part of the dissipated energy (about 1\%) is injected and damped through the magnetospheric twist decay, and 
an even smaller part reaches the surface through 
heat
diffusion.
It is through these two channels that the outburst produces
the
observable X-ray luminosity.


\section{Discussion}\label{discussion}

Hall evolution of the magnetic field provides a robust mechanism for growing 
magnetic stress in the solid crust of a magnetar, and yielding to theses stresses results 
in mechanical heating of the crust. \citet{2016arXiv160509077B} showed that heating from 
internally fostered crustal failures obeys strong constraints, which prevent it from sustaining 
the observed high temperatures of persistently luminous magnetars. In agreement with these 
constraints, our results do not show a strong persistent heating of magnetars. However,
the described mechanism of activity driven by Hall drift can have significant 
observational implications. The intermittent shear motions of the failed 
crust play a key role for twists of the external magnetosphere and also provide 
sporadic mechanical heating. This may explain outbursts of activity, in particular in 
the increasing number of so-called transient magnetars.

The model we have studied is one-dimensional, and therefore can only serve as a proxy for
the evolution of magnetic fields and crustal deformations of real three-dimensional magnetars. 
However, we expect
that the main features of the model, i.e. the avalanches of mechanical failures mediated by short wavelength Hall waves, large-scale thermoplastic waves, and the magnetospheric twists that these  cause, will be present in a more realistic three-dimensional dynamics. We note also that multidimensional simulations will likely show the interaction of crustal failures with the non-linear Hall dynamics that is not captured in our 1D model.
Magnetic instabilities in 2D or 3D, e.g. density-shear instability  \citep{2014PhPl...21e2110W,2015MNRAS.453L..93G}, will provide other channels to launch Hall waves by creating current 
sheets in the crust \citep{2016PNAS..113.3944G}.
In what follows we comment on how the simulated outbursts in our model compare with the observed 
outbursts.

During an outburst of a transient magnetar the observed X-ray flux increases by a factor 
of 10-1000 compared to its quiescent level and typically decays on the time scale of 
months to years. A canonical example is provided by
the first discovered  transient magnetar  XTE J1810-197 \citep{2004ApJ...609L..21I}, with the
characteristic dipole magnetic field $\sim 3\times 10^{14}$~G. 
It was discovered in January 2003 when its X-ray luminosity was comparable 
to $10^{35}$~erg~s$^{-1}$, which is a factor of $\sim 100$ above the quiescent level. 
It returned back to the quiescent level in a few years
\citep{2007Ap&SS.308...79G}.
However, no data is available for the early phase of this event (from Nov. 2002 to Jan. 2003) and so one cannot observationally study the rise of 
its light curve. 
The spectral fits of the outburst showed the appearance and subsequent shrinking of a 
hot spot on the star of size $\simlt 3$~km, which indicates a localized twist of the external 
magnetosphere. 
The transient magnetar discovered recently in the Galactic Centre SGR~1745$-$2900 \citep{2013ApJ...770L..24K,2013ApJ...770L..23M} is almost a twin of  XTE~J1810-197. Its outburst 
showed a similar decay, with a similar shrinking hot spot \citep{2014ApJ...786...84K,2015MNRAS.449.2685C}.
Similar strong outbursts were observed in several other transient magnetars 
(see \citet{2011ASSP...21..247R} for a review).

Less dramatic outbursts are also observed in ``persistent'' magnetars that show 
a continuously high level of emission during decades of observations. For instance,
the long-term observations of 1E 1048.1-5937 captured four outbursts 
(and resolved their rise times)
 between 2001 and 2007 \citep{2008ApJ...677..503T}. The 2001-2002 event 
increased the X-ray luminosity by a factor of $\sim 2$ over
$\sim 20$ days and then decayed over
$\sim$100 days \citep{2004ApJ...609L..67G}. The rise 
times of the
2002 and 2004 outbursts 
were a few weeks. The 2007 outburst rose to 
its peak in less than a week \citep{2008ApJ...677..503T}.

How does this data stack up against our model?  
The model predicts spikes in surface radiation flux of
$10^{22}-10^{24}$~erg~s$^{-1}$~cm$^{-2}$. Assuming an emission area of
about $10^{11}\hbox{cm}^2$ ($3\hbox{~km}\times 3\hbox{~km}$), our simulated 
peaks of luminosity are $10^{33}-10^{35}$~erg/s. 
The typical decay time of luminosity after the end of the failure event is typically
comparable to one year. These values are in good agreement with observations.

In our model, we have three different timescales: the timescale of thermoplastic waves 
(controlled by parameter $\alpha$), Hall-mediated avalanches, and heat diffusion. 
The heat diffusion timescale is  comparable to one year
and independent of the plastic-flow constant $\alpha$. 
It controls both the rise and the decay of surface luminosity due to 
heat diffusion from the heated interior to the surface. 
The observed rise times in the sources described above are 
often much shorter, more consistent with the timescale of magnetospheric twisting by 
thermoplastic waves. This suggests that the onset of the outburst is controlled by 
magnetospheric dissipation induced by the plastic motions of the crust. These motions
extract energy from the stellar interior (through Poynting flux) much faster than heat 
diffusion, and with a higher efficiency. A significant fraction of energy dissipated in the 
magnetosphere should be delivered to the surface by accelerated particles and 
radiated from the surface. There is strong observational evidence for this external heating 
of the magnetar surface, see \citet{2016arXiv160509077B}.

Our simulation shows that
the outburst rise time depends on whether the crustal failure 
develops through a Hall-mediated avalanche or a large-scale thermoplastic wave.
The rate of crustal failure (and the corresponding surface shear rate) in both cases is proportional 
to $\alpha^{1/2}$, see \Eqs~(\ref{eq:vTPW}) and (\ref{eq:vH}), and their ratio is independent 
of $\alpha$. The Hall-mediated avalanche is slower by the factor of $(D_{\rm H}/\chi)^{1/2}$,
where $D_{\rm H}=(cB_z/4\pi en_e)$ is the Hall diffusion coefficient and 
$\chi=\kappa/C_V\sim 10-100$~cm$^2$~s$^{-1}$ is the heat diffusion coefficient.
The factor $(D_{\rm H}/\chi)^{1/2}$ is typically around $10^{-2}$. The value of $\alpha$ is 
unknown, and both failure modes can be fast, giving short outburst rise times.
For the choice of parameters in our simulations, $\alpha=10^{-4}$~s$^{-1}$,
the typical outburst light-curve from a  thermoplastic wave rises to its peak 
in days to weeks.
The decay occurs on the much longer timescales of resistive magnetospheric untwisting
and heat diffusion through the crust. Both of these timescales are known to be roughly 
comparable to one year.

In our simulation, we see a large outburst every several hundred years. 
However, our simple 1D model simulates only a small patch on the magnetar surface
--- our simulation box may represent a crustal plate with surface area of a few
square kilometers (as the crust thickness is about one kilometer). 
There may be hundreds of such
independent patches, each undergoing its own series of outbursts. 
2D or 3D simulations will be required to model the global picture, which can give much 
more frequent outbursts. 
The outburst rate also increases with increasing $B_{\rm core}$.
Our simulations assumed $B_{\rm core}=6\times 10^{15}$~G, and a higher value would
increase the magnetic energy flux from the core into the crust and make it easier to initiate 
plastic failures in the deeper crust.

In this
paper we concentrated on the relatively slow dynamics of outbursts. Therefore our 
results do not 
directly
apply to the distinct class of magnetar bursts and flares that have much 
shorter durations, with rise times much shorter than one second. 
\citet{1995MNRAS.275..255T,1996ApJ...473..322T} proposed that the bursts result from sudden
``brittle" failures in the crust. 
It is, however, unclear how the compressed magnetized material with pressure 
well above the Coulomb lattice energy could be brittle 
\citep{2003ApJ...595..342J,2012MNRAS.427.1574L,2014ApJ...794L..24B}.
Therefore, it appears more likely that the fast flares result from explosive relaxation of the 
twisted magnetosphere \citep{1995MNRAS.275..255T,2006MNRAS.367.1594L,2012ApJ...754L..12P,2013ApJ...774...92P}. 
These magnetospheric explosions also produce sudden deformations of the crust 
\citep{2015ApJ...815...25L} which leave strong gradients in the crustal magnetic field 
and may be followed by accelerated Hall evolution. Both ``internal'' (brittle) and 
``external'' (magnetospheric) models 
could be related to the clusters of 
``storm bursts''  \citep{2006A&A...445..313G,2008ApJ...685.1114I,2010MNRAS.408.1387I,2011ApJ...739...94S}
if the Hall evolution induced by a burst leads to more bursts.

The avalanches of thermoplastic failures 
and heating of the crust may  
affect the rotation rate of the magnetar by changing 
the rotation of the neutron superfluid in the lower crust. The superfluid vortices could become 
unpinned from the crustal lattice, resulting in
timing anomalies ---
glitches or anti-glitches associated with outbursts.
We defer the study of this possibility to future work.

\acknowledgments
This work was supported by a Monash Research Acceleration grant, NASA grant NNX13AI34G,  and a grant from the Simons Foundation (\#446228, Andrei Beloborodov).

\bibliography{hall1D}

\clearpage

\end{document}